\pdfoutput=1
\documentclass[showpacs,preprintnumbers,amsmath,nofootinbib,amssymb]{revtex4}
\usepackage{epsfig}
\usepackage{dcolumn}
\usepackage{bm}
\DeclareGraphicsExtensions{.eps,.ps,.eps.gz,.ps.gz,.eps.Z,{}}

\newcommand{\aap}{{\em Astr. \& Astrophys. \/}}

\newcommand{\amm}{{\em Amer. Math. Monthly\/}} 

\newcommand{\mnras}{{\em Mon. Not. R. Astr. Soc. \/}}

\renewcommand{\prd}{{\em Phys. Rev. D \/}}
\newcommand{\physrep}{{\em Phys. Rep. \/}}

\newcommand{\vx}{{\bf x}}

\newcommand{\vd}{{\bf d}}

\newcommand{\vq}{{\bf q}}

\renewcommand{\vr}{{\bf r}}

\newcommand{\vk}{{\bf k}}
\newcommand{\ii}{{\rm i}}
\newcommand{\dd}{{\rm d}}

\newcommand{\mg}{\big<}
\newcommand{\md}{\big>}

\newcommand{\mF}{{\cal F}}
\newcommand{\mH}{{\cal H}}

\newcommand{\mM}{{\cal M}}

\newcommand{\mO}{{\cal O}}
\newcommand{\mP}{{\cal P}}

\newcommand{\mS}{{\cal S}}
\newcommand{\mR}{{\cal R}}
\newcommand{\mU}{{\cal U}}

\newcommand{\mT}{{\cal T}}

\newcommand{\Dirac}{\delta_{\rm D}}

\newcommand{\displ}{{\rm displ.}}
\newcommand{\tv}{\tau}

\newcommand{\beq}{\begin{equation}}
\newcommand{\eeq}{\end{equation}}
\newcommand{\beqa}{\begin{eqnarray}}
\newcommand{\eeqa}{\end{eqnarray}}

\def\la{\mathrel{\mathpalette\fun <}}

\def\fun#1#2{\lower3.6pt\vbox{\baselineskip0pt\lineskip.9pt
        \ialign{$\mathsurround=0pt#1\hfill##\hfil$\crcr#2\crcr\sim\crcr}}}

\def\Mpc{\, h^{-1} \, {\rm Mpc}}

\newcommand{\bdm}{\begin{displaymath}}
\newcommand{\edm}{\end{displaymath}}
\newcommand{\bea}{\begin{eqnarray}}
\newcommand{\eea}{\end{eqnarray}}
\newcommand{\bt}{\begin{tabular}}
\newcommand{\et}{\end{tabular}}
\newcommand{\xv}{{\bf x}}
\newcommand{\kv}{{\bf k}}

\newcommand{\qv}{{\bf q}}
\newcommand{\pv}{{\bf p}}
\newcommand{\de}{{\rm d}}

\def\d{\delta}

\def\Mpc{\, h^{-1} \, {\rm Mpc}}

\def\kMpc{\, h \, {\rm Mpc}^{-1}}

\def\fNL{f_{NL}}
\def\gNL{g_{NL}}

\def\la{\langle}
\def\ra{\rangle}
\def\ltsima{$\; \buildrel < \over \sim \;$}   
\def\gtsima{$\; \buildrel > \over \sim \;$} 
\def\simlt{\lower.5ex\hbox{\ltsima}}   
\def\simgt{\lower.5ex\hbox{\gtsima}}  

\begin{document}

\title{Multi-Point Propagators for Non-Gaussian initial conditions}
\author{Francis Bernardeau$^{1}$, Mart\'{\i}n Crocce$^{2}$ \& Emiliano Sefusatti$^{1}$}
\affiliation{$^{1}$Institut de Physique Th{\'e}orique,
         CEA/DSM/IPhT, Unit{\'e} de recherche associ{\'e}e au CNRS, CEA/Saclay
         91191 Gif-sur-Yvette c{\'e}dex}
\affiliation{$^{2}$Institut de Ci\`encies de l'Espai, IEEC-CSIC, Campus UAB,
Facultat de Ci\`encies, Torre C5 par-2,  Barcelona 08193, Spain}
\vspace{.2 cm}
\date{\today}
\vspace{.2 cm}
\begin{abstract}

We show here how Renormalized Perturbation Theory (RPT) calculations
applied to the quasi-linear growth of the large scale structure can be
carried on in presence of primordial non-Gaussian (PNG) initial
conditions. It is explicitly demonstrated that the series reordering
scheme proposed in Bernardeau, Crocce and Scoccimarro (2008) is
preserved for non-Gaussian initial conditions. This scheme applies to
the power spectrum and higher order spectra and is based on a
reorganization of the contributing terms into sum of products of
multi-point propagators. In case of PNG new contributing terms appear,
the importance of which is discussed in the context of current PNG models. 
The properties of the building blocks of such resummation schemes, the
multi-point propagators, are then investigated. It is first remarked
that their expressions are left unchanged at one-loop order
irrespectively of statistical properties of the initial field. We
furthermore show that the high-momemtum limit of each of these
propagators can be explicitly computed even for arbitrary initial conditions. They are found to be damped by an exponential cutoff whose expression is directly related to the moment generating function of the one-dimensional displacement field. This extends what had been established for multi-point propagators for Gaussian initial conditions. Numerical forms of the cut-off are shown for the so-called local model of PNG.
\end{abstract}
 \vskip2pc

\maketitle

\section{Introduction}

With the advent of precision measurements of the large-scale structure properties of the universe (e.g. \citep{2001MNRAS.327.1297P,2010MNRAS.401.2148P,2009MNRAS.400.1643S}), it became important to develop  semi-analytic tools that allow to accurately compute the large scale statistical properties of the cosmic fields
for any cosmological model. Indeed, it has been realized that the
simple linear theory is too crude for the precision one wishes to
attain. This is the case for instance for the precise determination of
the power spectrum at the scales of the baryonic acoustic oscillations (see for instance \cite{2008PhRvD..77b3533C,2010MNRAS.404...60R,2008MPLA...23...25M}). 

The Standard Perturbation Theory approach (see
\cite{2002PhR...367....1B}) is  however not very efficient in
producing well behaved next-to-leading order terms to the spectra or
bispectra. It leads to perturbative series that have poor convergence
properties (see \cite{2006PhRvD..73f3519C} for more insights).  Various resummation schemes have been proposed over the last few years to overcome those difficulties while aiming at providing tools for performing such calculations in a systematic way \cite{2006PhRvD..73f3519C,2006PhRvD..73f3520C,2007astro.ph..3563M,2008JCAP...10..036P,2008ApJ...674..617T,2008PhRvD..78j3521B}. We will be
interested here more particularly in the so called Renormalized
Perturbation Theory (RPT) approach developed initially in
\cite{2006PhRvD..73f3519C} and \cite{2008PhRvD..78j3521B} where
rewritings of the resummation series are proposed. Whereas RPT as
presented in \cite{2006PhRvD..73f3519C} makes use of the one-point
propagator only, in \cite{2008PhRvD..78j3521B} it was introduced the concept of multi-point propagators that allow an alternative scheme for the computation of spectra and bispectra. 

Such developments however, have been so far limited to cases
corresponding to Gaussian initial conditions (but see \cite{2010JCAP...03..011B} for a
scheme that numerically solves the truncated integro-differential
equations governing the dynamics in presence of PNG) and it should be stressed that this is not only for convenience. In a diagrammatic representation of the dynamics, as presented for instance in \cite{2002PhR...367....1B} or in \cite{2006PhRvD..73f3519C} 
the assumptions on the statistical properties of the initial field are indeed crucial. At the same time, new ideas to constrain the non-Gaussian properties of the initial metric fluctuations from observations of the large-scale structure of the universe have emerged (see for instance the recent reviews \cite{2010arXiv1001.4707L,2010arXiv1001.5217V} and \cite{2010arXiv1003.5020D}). They suggest that future redshift surveys could be our best opportunity to probe a departure from Gaussian initial conditions. This is in fact our main motivation for the
investigations presented in this paper, where we extend the schemes and
results first introduced in \cite{2008PhRvD..78j3521B} to the case of primordial non-Gaussianities (PNG). The findings presented here are rather general and, except for some numerical illustrations, are valid for any types of non-Gaussian initial conditions.

The paper is organized as follows. In Sec.~\ref{sec:motionequations}
we recall the equations of motion for gravitational instabilities of a
cosmic fluid. In Sec.~\ref{sec:gammaexpansion} we extend the expansion of power spectrum and bispectrum in terms of multi-point
 propagators to the case of arbitrary initial statistics. In Sec~\ref{sec:largeklimit}
we compute explicitly the multi-point propagators in the large-$k$ limit
for non Gaussian initial conditions. Lastly, in Sec.~\ref{sec:localmodel} we study
the concrete example of primordial non Gaussianities of the local
type. Our concluding remarks are presented in Sec.~\ref{sec:conclusions}.

\section{The equations of motion}
\label{sec:motionequations}

We are interested here in the development of cosmological instabilities in a cosmological dust fluid. In general the dynamical evolution of such a fluid can be described with the Vlasov equation. As usual we restrict our investigations to the regime where multi-flow regions play a negligible role. In the one flow limit, the equations of motion then take the form of a set of three coupled equations relating the local density contrast, the peculiar velocity field and the gravitational potential (see \cite{2002PhR...367....1B}). 

At {\it linear order} these
equations can easily be solved for an arbitrary background cosmology.
One generically finds a growing solution and a decaying solution in the
density contrast and peculiar velocity fields satisfying,
\begin{eqnarray}
\frac{\partial}{\partial \tau}\delta(\vk,\tau)+\theta(\vk,\tau)&=&0\,, \label{massconservation}\\
\frac{1}{\mH}\frac{\partial}{\partial
  \tau}\frac{\theta(\vk,\tau)}{\mH}+\left(1+\frac{1}{\mH^2}\frac{\partial\mH}{\partial
    \tau}\right)\frac{\theta(\vk,\tau)}{\mH}+\frac{3}{2}\,\Omega_{m}(\tau)\delta(\vk,\tau)&=&0\,,
\label{momentumconservation}
\end{eqnarray}
 with $\delta(\vk,\tau)$ and $\theta(\vk,\tau)$ being respectively the Fourier transforms of the density contrast,  $\delta (\vx,\tau)=\rho(\vx)/\bar \rho - 1$, and of the peculiar velocity divergence, $\theta\equiv\nabla\cdot{\bf v}$. In Eqs.~(\ref{massconservation},\ref{momentumconservation}), $\Omega_m(\tau)$ is the matter density and ${\cal H}\equiv {d\ln a
/{d\tau}}$  is the conformal expansion rate  with  $a(\tau)$ the cosmological scale factor and
$\tau$ the conformal time. 
If one denotes $D_{+}(\tau)$ the growing mode solution of this system and
$f_{+}(\tau)$ its logarithmic derivative with respect to the expansion
then,
\begin{equation}
\delta(\vk,\eta)=D_{+}(\eta)\delta_{0}(\vk)\,,\ \ \theta(\vk,\eta)/\mH=-f_{+}(\eta)D_{+}(\eta)\delta_{0}(\vk)\,,
\end{equation}
is the solution for the growing mode and 
\begin{equation}
\delta(\vk,\eta)=D_{-}(\eta)\delta_{0}(\vk)\,,\ \ \theta(\vk,\eta)/\mH=-f_{-}(\eta)D_{-}(\eta)\delta_{0}(\vk)\,,
\end{equation}
for the decaying. We can note that in the Einstein-de Sitter limit we
have $f_{+}=1$ and $f_{-}=-3/2$, leading to $D_+=a$ and
$D_-=a^{-3/2}$. However, in the more generic case of arbitrary matter
and dark-energy content $f_{+}$ and $f_{-}$ do not admit an analytic
solution \cite{2002PhR...367....1B}.

Following \cite{1998MNRAS.299.1097S}, the equations of motion can be written
in a compact form with the use of the two component quantity $\Psi_i(\vk,\tau)$, defined as 
\begin{equation}
\Psi_i(\vk,\tau) \equiv \Big( \delta(\vk,\tau),\ -\frac{1}{f_{+}(\tau)\cal H}\theta(\vk,\tau) \Big),
\label{2vector}
\end{equation}
where the index $i=1,2$ selects the density or velocity components and which makes explicit use of the growing solution. It is then convenient to re-express the time-dependence in terms of the growing solution and in the following we will use the time variable $\eta$
defined as 
\begin{equation}
\eta=\log {D_{+}(\eta)}\,,
\end{equation}
assuming the growing factor set to unity at initial time. Then the {\it fully nonlinear} equations in Fourier space read \cite{2002PhR...367....1B} (we henceforth use the convention that repeated Fourier arguments are integrated over),
\begin{equation}
\frac{\partial}{\partial \eta} \Psi_i(\vk,\eta) + \Omega_{ij}(\eta) \Psi_j(\vk,\eta) =
\gamma_{ijk}(\vk,\vk_1,\vk_2) \ \Psi_j(\vk_1,\eta) \ \Psi_k(\vk_2,\eta),
\label{eom}
\end{equation}
where 
\begin{equation}
\Omega_{ij} (\eta) \equiv \Bigg[ 
\begin{array}{cc}
0 & -1 \\ -\frac{3}{2}\frac{\Omega_{m}}{f_{+}^2} & \frac{3}{2}\frac{\Omega_{m}}{f_{+}^2}-1 
\end{array}        \Bigg],
\end{equation}
and the {\sl symmetrized vertex} matrix $\gamma_{ijk}$ describes the non linear 
interactions between different Fourier modes. Its components are given by
\begin{eqnarray}
\gamma_{222}(\vk,\vk_1,\vk_2)&=&\Dirac(\vk-\vk_1-\vk_2) \ {|\vk_1+\vk_2|^2 (\vk_1
\cdot\vk_2 )\over{2 k_1^2 k_2^2}}, \nonumber \\
\gamma_{121}(\vk,\vk_1,\vk_2)&=&\Dirac(\vk-\vk_1-\vk_2) \  {(\vk_1+\vk_2) \cdot
\vk_1\over{2 k_1^2}},
\label{vertexdefinition}
\end{eqnarray}
with $\gamma_{ijk}(\vk,\vk_a,\vk_b)=\gamma_{ikj}(\vk,\vk_b,\vk_a)$, and $\gamma=0$ 
otherwise, where $\Dirac$ denotes the Dirac delta distribution. The matrix  $\gamma_{ijk}$ is independent 
on time (and on the background evolution) and encodes all the
non-linear couplings of the system.
 The formal integral solution to Eq.~(\ref{eom}) is given by (see \cite{1998MNRAS.299.1097S,2001NYASA.927...13S,2006PhRvD..73f3519C} for a detailed derivation) 
\begin{equation}
\Psi_i(\vk,\eta) = g_{ij}(\eta) \ \phi_j(\vk) +  \int_0^{\eta}  {\dd \eta'} \ g_{ij}(\eta,\eta')\, \gamma_{jkl}^{(\rm s)}(\vk,\vk_1,\vk_2)\, \Psi_k(\vk_1,\eta')\, \Psi_l(\vk_2,\eta'),
\label{eomi}
\end{equation}
where  $\phi_i(\vk)\equiv\Psi_i(\vk,\eta=0)$ denotes the initial conditions, set when the growth factor $D_{+}=1$ and where $g_{ij}(\eta)$ is the {\em linear propagator}, that is the Green's function of the linearized version of Eq.~(\ref{eom}) and describes the standard linear evolution of the density and velocity fields from their initial state.  It is such that $g_{ab}(\eta,\eta') =0$ for $\eta<\eta'$ due to causality, and $g_{ab}(\eta,\eta') \rightarrow \delta_{ab}$ as $\eta-\eta'\rightarrow 0^{+}$. It naturally
encapsulates the linear solutions growth and reads
\begin{equation}
g_{ij}(\eta,0) = \frac{D_{+}(\eta)}{5}
\Bigg[ \begin{array}{rr} 3 & 2 \\ 3 & 2 \end{array} \Bigg] -
\frac{D_{-}(\eta)}{5}
\Bigg[ \begin{array}{rr} -2 & 2 \\ -2\frac{f_{-}(\eta)}{f_{+}(\eta)} & 2\frac{f_{-}(\eta)}{f_{+}(\eta)} \end{array} \Bigg],
\label{prop}
\end{equation} 
for $\eta\geq 0$ assuming the initial conditions (for $a=1$) are set at a time when the universe is very close to Einstein-de Sitter.
We note that growing mode initial conditions
correspond to setting $\phi_{1}(\vk)=\phi_{2}(\vk)$, in such a way that the second term in Eq.~(\ref{prop}) does not intervene.
The general expression (see \cite{2008JCAP...10..036P}) of the propagator $g_{ij}(\eta,\eta')$ is obtained from the property
\begin{equation}
g_{ij}(\eta,\eta')g_{jk}(\eta',0)=g_{ik}(\eta,0)\,,\label{gcomp}
\end{equation}
so that
\begin{equation}
g_{ij}(\eta,\eta')=g_{ik}(\eta,0)g^{-1}_{kj}(\eta',0).
\end{equation}
We can remark that for a Einstein-de Sitter background $g_{ij}(\eta,\eta')=g_{ij}(\eta-\eta',0)$.

\section{Statistics with Non-Gaussian initial conditions}  
\label{sec:gammaexpansion}

\subsection{The $\Gamma-$expansion}
\label{subsec:gammaexpansion}

We are interested in the statistical properties of the density and
velocity divergence fields, in particular in the construction of their power spectra and higher-order correlators. The (equal time) power spectra $P_{ij}$ are defined as
\begin{equation}
\langle \Psi_{i}(\vk)\Psi_{j}(\vk')\rangle=\Dirac(\vk+\vk')P_{ij}(k)\,,
\end{equation}
and we want to reconstruct them from the expression of $\Psi_{i}(\vk)$ in terms of the initial density field $\phi_{j}(\vk)$, the correlation properties of which are assumed to be known. The aim of this paper is precisely to explore the effects of dropping the assumption that the initial conditions are Gaussian distributed.

A perturbative solution to Eq.~(\ref{eomi}) can be obtained by expanding the fields in terms of the initial ones,
\begin{equation}
\Psi_{i}(\vk,\eta)=\sum_{n=1}^{\infty}\Psi_{i}^{(n)}(\vk,\eta)\,,
\label{PsiExpansion}
\end{equation}
such that,
\begin{eqnarray}
\Psi_{i}^{(n)}(\vq,\eta)=\int\dd^3\vq_{1}\dots\dd^3\vq_{n}\ 
\Dirac(\vq-\vq_{1\dots n}) \,\mF^{(n)}_{i j_1 j_2 \ldots j_n}(\vq_{1},\dots,\vq_{n};\eta)\ 
\phi_{j_1}(\vq_{1})\dots\phi_{j_n}(\vq_{n})\,,
\label{mFndef}
\end{eqnarray}
where we adopt the notation $\vq_{1\dots n}\equiv\vq_1+\dots+\vq_n$ for vectors sums and where $\mF^{(n)}$ are fully symmetric functions of the wave-vectors that can be obtained recursively in terms of $g_{ij}$ and $\gamma_{ijk}$ \cite{2002PhR...367....1B}.  Note that these functions have a non-trivial time-dependence because they also include sub-leading terms in $\eta$. Their fastest growing contribution is of course given by the well known $\{F_n,G_n\}$ kernels in PT (assuming growing mode initial conditions) whose dependence on time we recall here for an Einstein-de Sitter background,
\begin{equation}
\mF^{(n)}_i=\exp{(n\eta)}\ \{F_n(\vq_1,..,\vq_n),G_n(\vq_1,..,\vq_n)\}\,,
\label{Fnkernels}
\end{equation}
for $i=1,2$ (density or velocity divergence fields respectively).

A formal expression for $P_{ij}(k)$ can be written using the expansion in Eq.~(\ref{PsiExpansion}) as,
\begin{eqnarray}
\Dirac(\vk_{1}+\vk_{2})P_{ij}(k_{1},\eta)
&=&\sum_{n_{1},n_{2}}\mg\Psi^{(n_{1})}_{i}(\vk_{1},\eta)\Psi^{(n_{2})}_{j}(\vk_{2},\eta)\md.
\label{PkExpansion1}
\end{eqnarray}
For simplicity in what follows we will simply drop the component
indices $i$ and $j$ but they are implicit. 
Then, for a given choice of indices $n_{1}$ and $n_{2}$ one has to
compute the ensemble average of $n_{1}+n_{2}$ factors $\phi(\vq_{i})$,
following the field expansion in Eq.~(\ref{mFndef}). The ensemble
average of $\phi(\vq_{1})\dots\phi(\vq_{n_{1}})
\phi(\vq_{n_{1}+1})\dots\phi(\vq_{n_{1}+n_{2}}) $ is given by the sum
of product of cumulants of all set of subsets of
$\{\phi(\vq_{1})\dots\phi(\vq_{n_{1}})
\phi(\vq_{n_{1}+1})\dots\phi(\vq_{n_{1}+n_{2}}) \}$ that form a
partition. Let us be more precise. Let us define $\mP_{n}$ the set of
partitions of a set of $n$ variables \footnote{There is no simple expression for the number of such partitions. Its number is called the Bell number, $B_{n}$, after E.T. Bell, \cite{Bell1934}.}. Its elements $\mS_{i}$ are lists of subsets, $s_{i}$, and each element of $s_{i}$ is an index in the $1\dots n$ range. As a result,
\begin{equation}
\mg\phi(\vq_{1})\dots\phi(\vq_{n_{1}}) \phi(\vq_{n_{1}+1})\dots\phi(\vq_{n_{1}+n_{2}}) \md=
\sum_{\mS_{i}\in\mP_{n_{1}+n_{2}}}\prod_{s_{i}\in\mS_{i}}\mg\prod_{i\in s_{i}}\phi(\vq_{i})\md_{c}\,,
\end{equation}
where $\mg\dots\md_{c}$ are the cumulants. Contrary to the case of Gaussian initial conditions, there exist cumulants that involve more than two variables. We can still assume though that there are no singleton although this hypothesis does not change the end of the calculation.

The idea is now to sort the elements of $\mP_{n_{1}+n_{2}}$. Each
element $\mS_{i}$ of $\mP_{n_{1}+n_{2}}$ defines four cardinal
numbers; the numbers $q_{1}$ and $q_{2}$ of points in subsets that are
entirely within the first $n_{1}$ points or the last $n_{2}$ points
and the numbers $p_{1}=n_{1}-q_{1}$ and $p_{2}=n_{2}-q_{2}$ of points
within the first $n_{1}$ or last $n_{2}$ that are in parts that are
neither a subset of $\{1,\dots, n_{1}\}$ nor of $\{n_{1}+1,\dots,
n_{2}\}$. The resulting value of the partition contribution to the
moment does not depend on which of the $p_{1}$ or $p_{2}$ points is
thus chosen because the $\mF^{n}$ functions are fully symmetric in
their arguments. Up to symmetry factors $ \binom{n_{1}}{p_{1}}
\binom{n_{2}}{p_{2}}$, it is then possible to assume that the $p_{1}$
points correspond to the first  ones of $n_{1}$ and similarly for
$p_{2}$. 

\begin{figure}[htbp]
\begin{center}
\includegraphics[width=0.9\textwidth]{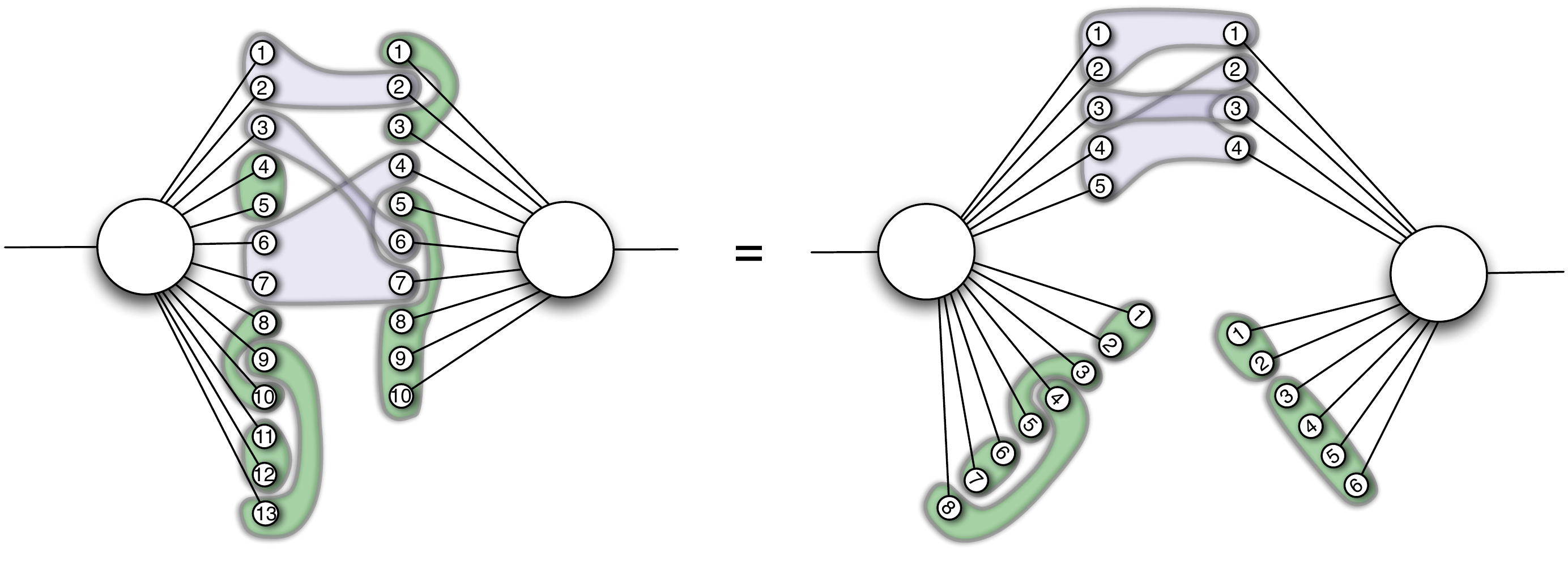}
\caption{Example of partitions of $n_{1}+n_{2}$ points (here $n_{1}=13$ and $n_{2}=10$). On each ensemble
the elements have been numbered. For this choice of partitions there
are 5 elements out of 13 that are in subsets that span both column and
4 out of 10; therefore $p_{1}=5$, $q_{1}=8$ and $p_{2}=4$,
$q_{2}=6$. The left hand side shows a generic partition of that
kind. The right panel shows a partition which is identical to the left
one up to a renumbering of the indices.  These  two contributing diagrams take the same values. When the order of the indices in each subset is preserved, the number of such contributing terms (e.g. the number of diagrams that lead to the same right hand side figure) is $ \binom{p_{1}+q_{1}}{p_{1}} \binom{p_{2}+q_{2}}{p_{2}}$. The summation over all possible partitions can then be restricted to those of the type of the left hand side with a weight given by the aforementioned symmetry factor. Such terms are defined by three new partitions that are elements respectively of
$\mP_{q_{1}}$, $\mP_{q_{2}}$ and $\mP^{X}_{p_{1},p_{2}}$ (see text for definitions). }
\label{GExpNGinit}
\end{center}\end{figure}

Let us now define $\mP_{q_{1},p_{1},q_{2},p_{2}}$ as a subset of $\mP_{n_{1}+n_{2}}$, e.g. the set of partitions in $\mP_{n_{1}+n_{2}}$
with fixed values of $q_{1}, p_{1}, q_{2}$ and $p_{2}$ that are formed from the union
of an element of $\mP_{q_{1}}$, an element of $\mP_{q_{2}}$ and an
element of $\mP^{X}_{p_{1},p_{2}}$. The latter is defined as the set
of all partitions of $\{1,\dots,p_{1}+p_{2}\}$ that do not contain
subsets contained entirely within the first $n_{1}$ points or within
the last $n_{2}$. In Fig.~\ref{GExpNGinit} we give an explicit example of
this partitioning scheme where an example of two different partitions having the same value are presented.

As a result, the power spectrum in Eq.~(\ref{PkExpansion1}) reads
\begin{eqnarray}
\Dirac(\vk_{12})P_{ij}(k_{1},\eta)\! &=&\!\sum_{n_{1},n_{2}}\!\int\!\dd^3\vq_{1}\dots\dd^3\vq_{n_{1}+n_2} 
\Dirac(\vk_1\!-\!\vq_{1\dots n_1}) \Dirac(\vk_2\!-\!\vq_{n_{1}+1\dots n_1+n_2}) \times 
\nonumber\\
& & 
\mF^{(n_{1})}(\vq_1,\dots,\vq_{n_1}) \mF^{(n_{2})}(\vq_{1+n_1},\dots,\vq_{n_1+n_2})\sum_{\mS_{i}\in\mP_{n_{1}+n_{2}}}\prod_{s_{i}\in\mS_{i}}\mg\prod_{i\in
  s_{i}}\phi(\vq_{i})\md_{c}\,,
  \label{expansion}
\end{eqnarray}
with
\begin{eqnarray}
   \sum_{\mS_{i}\in\mP_{n_{1}+n_{2}}}\prod_{s_{i}\in\mS_{i}}\mg\prod_{i\in
  s_{i}}\phi(\vq_{i})\md_{c}  
&=&
\sum_{q_{1},p_{1},q_{2},p_{2}} \binom{p_{1}+q_{1}}{p_{1}} \binom{p_{2}+q_{2}}{p_{2}} 
\sum_{\mS_{i}\in\mP_{q_{1},p_{1},q_{2},p_{2}}}\prod_{s_{i}\in\mS_{i}}\mg\prod_{i\in
  s_{i}}\phi(\vq_{i})\md_{c} \nonumber \\
&=&\sum_{q_{1},p_{1},q_{2},p_{2}}\ \binom{p_{1}+q_{1}}{p_{1}}
\binom{p_{2}+q_{2}}{p_{2}}\left(\sum_{\mR_{i}\in\mP_{q_{1}}}\prod_{r_{i}\in\mR_{i}}\mg\prod_{i\in r_{i}}\phi(\vq_{p_{1}+i})\md_{c}\right)  \nonumber\times\\
&&
\left(\sum_{\mT_{i}\in\mP_{q_{2}}}\prod_{t_{i}\in\mT_{i}}\mg\prod_{i\in t_{i}}\phi(\vq_{n_{1}+p_{2}+i})\md_{c}\right)
\left(\sum_{\mU_{i}\in\mP^{X}_{p_{1},p_{2}}}\prod_{u_{i}\in\mU_{i}}\mg\prod_{i\in u_{i}}\phi(\vr_{i})\md_{c}\right)\,, 
\end{eqnarray}
where we have introduced the set of wave numbers $\{\vr_{i}\}$ that corresponds to reindexation of some of the $\vq_{i}$,
\begin{equation}
\{\vr_{1},\dots,\vr_{q_{1}+q_{2}}\}\equiv\{\vq_{1},\dots,\vq_{q_{1}},\vq_{n_{1}+1},\dots,\vq_{n_{1}+q_{2}}\},
\end{equation}
and where $\sum_{q_{1},p_{1},q_{2},p_{2}}$ stands for
$\sum_{q_{1}=0}^{n_{1}}\sum_{p_{1}=0}^{n_{1}-q_{1}}\sum_{q_{2}=0}^{n_{2}}\sum_{p_{2}=0}^{n_{2}-q_{2}}$. The
crucial property is then that the first two parenthesis that appear in
the last line of the  previous expression
depend only on $q_{1}$ or $q_{2}$ respectively. The sum over
partitions can then be reorganized by summing for fixed values of
$p_{1}$ and $p_{2}$ first (see Fig.~\ref{GExpNGinit}, right panel). That is, by doing
\begin{displaymath}
\sum_{n\ge 0} \mF^{(n)}\sum_{q=0}^{n}\sum_{p=0}^{n-q}\binom{n}{p}\rightarrow
\sum_{p\ge 0} \sum_{q\ge 0}\binom{p+q}{q}\mF^{(p+q)}
\end{displaymath}
in each index $1$ and $2$ (after using that $n=p+q$) one can identify the function $\Gamma^{(p)}$ defined as,
\begin{eqnarray}
\Gamma^{(p)}(\vq_{1},\dots,\vq_{p},\eta)&=&
\sum_{q=0}^{\infty} \binom{p+q}{q} \int\dd^3\vq_{1}\dots\dd^3\vq_{q} \,
\mF^{(p+q)}(\vq_{1},\dots,\vq_{p},\vq_{p+1},\dots,\vq_{p+q};\eta) \nonumber \\ 
&&\times\,\left(\sum_{\mR_{i}\in\mP_{q}}\prod_{r_{i}\in\mR_{i}}
\ \mg\prod_{i\in r_{i}}\phi(\vq_{p+i})\md_{c}\right), 
\label{GpdefNG}
\end{eqnarray}
(here $i\in[1,q]$)
which naturally extends to the case of arbitrary initial statistics the results
for $\Gamma^{(p)}$ studied in
\cite{2006PhRvD..73f3520C,2008PhRvD..78j3521B} for Gaussian initial
conditions. In such case $q$ takes only even values and the two point
initial spectrum determines the partition $\mP_q$, see for instance Eq.~(15) in \cite{2006PhRvD..73f3520C}
and Eq.~(21) in \cite{2008PhRvD..78j3521B} for $\Gamma^{(1)}$ and $\Gamma^{(2)}$ respectively. 

After inserting Eq.~(\ref{GpdefNG}) back into Eq.~(\ref{expansion}) we arrive at,
\begin{eqnarray}
\Dirac(\vk_{12})P(k_{1},\eta)
&=&\sum_{p_{1},p_{2}}\ \int\dd^3\vq_{1}\dots\dd^3\vq_{p_{1}}
\int\dd^3\vq_{p_{1}+1}\dots\dd^3\vq_{p_{1} + p_{2}} 
\Dirac(\vk_1-\vq_{1\dots p_1}) \Dirac(\vk_2-\vq_{p_{1}+1 \dots
  p_1+p_2}) \nonumber \\
&& \Gamma^{(p_{1})}(\vq_{1},\dots,\vq_{p_{1}},\eta)
\Gamma^{(p_{2})}(\vq_{p_1+1},\dots,\vq_{p_1+p_2},\eta)
\sum_{\mU_{i}\in\mP^{X}_{p_{1},p_{2}}}\ 
\prod_{u_{i}\in\mU_{i}}\mg\prod_{i\in u_{i}}\phi(\vq_{i})\md_{c},
\label{PkExpansionNG}
\end{eqnarray}
which is the final expression for the $\Gamma$-expansion of the power spectrum.
It naturally extends the $\Gamma$-expansion obtained in \cite{2008PhRvD..78j3521B} for
Gaussian initial conditions to an arbitrary initial statistics. Note that contrary to that case the sum
is not restricted to $p_{1}=p_{2}$. As a consequence there is in
general no guarantee that all terms of this sum are positive. In addition,
note that from Eq.~(\ref{expansion}) to Eq.~(\ref{PkExpansionNG})
$\Dirac(\vk-\vq_{1\dots p})$ could automatically be factorized out
since $\mg\prod_{i\in r_{i}}\phi(\vq_{p+i})\md_{c}$ are all
proportional to $\Dirac(\vq_{p+1\dots p+q})$.

\begin{figure}[htbp]
\begin{center}
\includegraphics[width=0.8\textwidth]{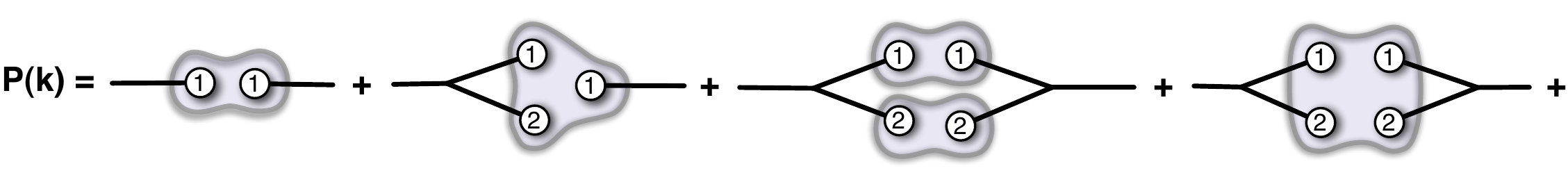} 
\caption{$\Gamma$-expansion for the power spectrum up to one-loop. The shaded regions indicate the initial (or
  ``primordial'') statistics corresponding to each diagram, that is: one initial power spectra (2-points) in the left most diagram, one
  initial bispectrum in the center-left one, two power spectra in the  center-right and one initial trispectrum in the
  right most figure.}
\label{fig:Pkexpansion}
\end{center}\end{figure}

In Fig.~\ref{fig:Pkexpansion} we show this expansion diagrammatically
up to one-loop terms. Explicitly, these diagrams corresponds to

\begin{eqnarray}
P(k)&=&[\Gamma^{(1)}(k)]^2\,P_0(k)+  2 \,\Gamma^{(1)}(k) \,\int \vd^3q \,
\Gamma^{(2)}(\vq,\vk-\vq) B_0(\vq,\vk-\vq,-\vk)  +  \nonumber \\
&& + \, 2 \int d^3q[\Gamma^{(2)}(\vq,\vk-\vq)]^2 P_0(\vk-\vq) P_0(\vq) + \nonumber \\
&& + \int \vd^3q_1
\vd^3q_2\,\Gamma^{(2)}(\vq_1,\vk-\vq_1)\,\Gamma^{(2)}(\vq_2,-\vk-\vq_2)\,T_0(\vq_1,\vk-\vq_1,\vq_2,-\vk-\vq_2)\,. 
\label{pkexplicit}
\end{eqnarray}

We stress that each term in the $\Gamma$-expansion above involves an
infinite number of perturbative contributions in $\delta$ as the
$\Gamma^{(p)}$ functions correspond to a full resummation of the propagator {\it and} the vertex in the language of \cite{2008PhRvD..77b3533C,2007astro.ph..3563M}.

\begin{figure}[htbp]
\begin{center}
\includegraphics[width=0.9\textwidth]{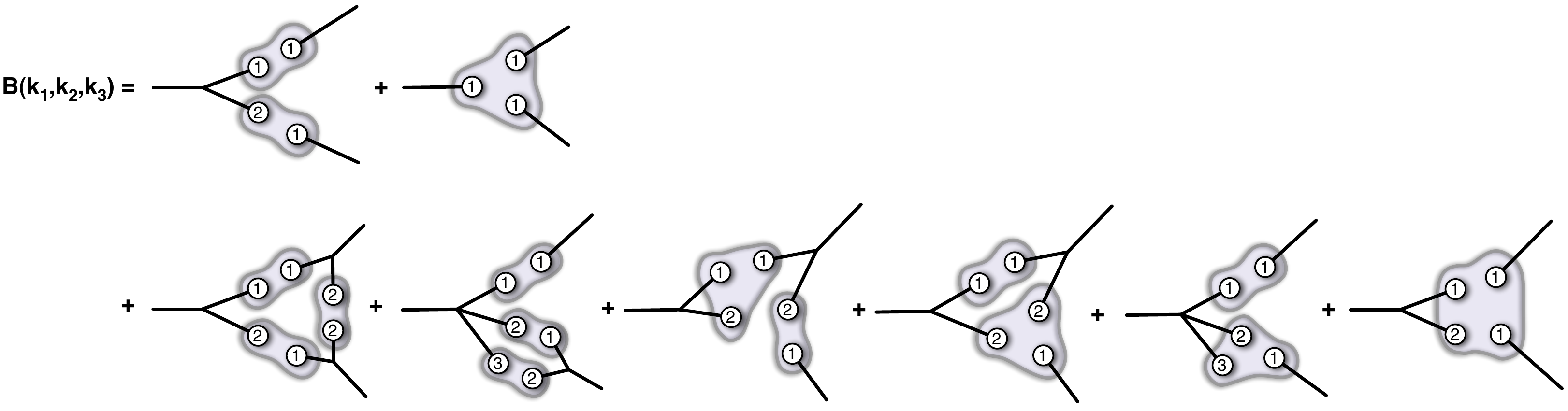}
\caption{$\Gamma$-expansion for the bispectrum up to one-loop. The
  first two contributions are proportional to
  $P_{0}^2$ (``tree level''), the remaining ones to $P_{0}^3$, assuming that the bispectrum $B_{0}$ and the trispectrum $T_{0}$ scale respectively as $P_{0}^2$ and $P_{0}^3$.
  The meaning of shaded regions is described in the caption
  to Fig.~\ref{fig:Pkexpansion} (see text for details).}
\label{fig:Bkexpansion}
\end{center}\end{figure}

The resummation leading to the $\Gamma$-expansion for $P(k)$ can
obviously be extended for higher order correlators: higher order
multi-point spectra can be obtained by gluing together $\Gamma^{(p)}$
functions multiplied by the proper cumulant. In the case of 3-point
statistics (bispectrum) there are $3$ ``external'' legs and the
equivalent of Eq.~(\ref{PkExpansion1}) for the bispectrum
$B(\vk_1,\vk_2,\vk_3)$ now runs over indices $n_1$, $n_2$ and
$n_3$. Following the same line of reasoning as above, the total set
of partitions $\mP_{n_1+n_2+n_3}$ is broken up into
three subsets $\mP_{q_i}$ contained fully with $n_i$ and one with ``cross''
elements $\mP^X_{p_1, p_2, p_3}$. The bispectrum is then a sum over the
product of elements of $\mP^X_{p_1, p_2, p_3}$ times
$\Gamma^{(p_1)}\,\Gamma^{(p_2)}\,\Gamma^{(p_3)}$, as in Eq.~(\ref{PkExpansionNG}). Figure~\ref{fig:Bkexpansion} shows all the contributions in the multi-point propagator
expansion of $B(\vk_1,\vk_2,\vk_3)$ up to one-loop diagrams. For PNG models satisfying the hierarchical scaling $B_0\sim P_0^2$, $T_0\sim P_0^3$ and so on these are all the terms up to $\mathcal{O}(P_0^3)$. These diagrams yield the following explicit expressions,

\begin{eqnarray}
B(k_1,k_2,k_3)&=&  2 \, \Gamma^{(2)}(\vk_1,\vk_2) \Gamma^{(1)}(k_1)
\Gamma^{(1)}(k_2) P_0(k_1)  P_0(k_2) + 2 \, {\rm perm.} + 
\nonumber \\ && 
\Gamma^{(1)}(k_1) \Gamma^{(1)}(k_2) \Gamma^{(1)}(k_3)
\,B_0(k_1,k_2,k_3) \ + \nonumber \\ 
&& 8\! \int\! d^3q\, \Gamma^{(2)}(-\vq,\vq\!+\!\vk_1)\,
\Gamma^{(2)}(-\vq\!-\!\vk_1,\vq\!-\!\vk_2)\, \Gamma^{(2)}(\vk_2\!-\!\vq,\vq)\,
P_0(q)\,P_0(|\vk_1\!+\!\vq|)\, P_0(|\vk_2\!-\!\vq|) + \nonumber \\
&& 6\, P_0(k_1)\, \Gamma^{(1)}(k_1) \!\int\! d^3q \,\Gamma^{(3)}(\vk_1,\vk_2\!-\!\vq,\vq)\,
\Gamma^{(2)}(\vk_2\!-\!\vq,\vq)\, P_0(q)\, P_0(|\vk_2-\vq|) +  5 \, {\rm
  perm.} \nonumber \\
&& 2\, P_0(k_1)\, \Gamma^{(1)}(k_1) \left[ \Gamma^{(2)}(\vk_1,\vk_2)\! \int \!d^3q\, \Gamma^{(2)}(\vq,\vk_2\!-\!\vq)\,
B_0(k_2,q,|\vk_2\!-\!\vq|) + (k_2 \leftrightarrow k_3) \right] + 2 \, {\rm
  perm.} \nonumber \\
&& 4 \, \Gamma^{(1)}(k_1)\! \int\! d^3q\, 
\Gamma^{(2)}(\vq,\vk_2\!-\!\vq)\, \Gamma^{(2)}(\vk_1\!+\!\vq,\vk_2\!-\!\vq)\, P_0(|\vk_2\!-\!\vq|)\,
B_0(k_1,q,|\vk_1\!+\!\vq|) + 2 \, {\rm perm.} + \nonumber \\
&& 3 \, \Gamma^{(1)}(k_1) \,\Gamma^{(1)}(k_2) \,P_0(k_1)\! \int\! d^3q\, 
\Gamma^{(3)}(\vk_1,\vq,\vk_2\!-\!\vq)\, B_0(k_2,q,|\vk_2\!-\!\vq|) + (k_1
\leftrightarrow k_2) + 2 \, {\rm perm.} + \nonumber \\
&& \Gamma^{(1)}(k_1)\, \Gamma^{(1)}(k_2)\!  \int\!
d^3q\,\Gamma^{(2)}(\vq,\vk_3\!-\!\vq)\, T_0(\vk_1,\vk_2,\vq,\vk_3\!-\!\vq) + 2 \, {\rm perm.}\,,
\label{bkexplicit}
\end{eqnarray}
where in the r.h.s. we implicitly assumed $\vk_3=-\vk_1-\vk_2$ (and so the l.h.s only depends on the magnitudes of the wave-vectors). The first two contributions are ``tree level''. The next two correspond to the gravitationally induced non-Gaussianity, while the last four arise from PNG.

Comparing with the standard perturbative approach up to one-loop given in \cite{2009PhRvD..80l3002S}
we see that his terms $B^{I}_{112}$, $B_{111}$, $B^{I}_{222}$,
$B^{I}_{123}$, $B^{I}_{122}$, $B^{II}_{122}$, $B^{II}_{113}$, $B^{II}_{112}$ correspond to
those in Eq.~(\ref{bkexplicit}) when the $\Gamma^{(p)}$ functions
are taken at their lowest or ``bare'' order in Eq.~(\ref{GpdefNG}) (i.e. prior to
any resummation). In turn, his $B^{II}_{123}$ ($B^{I}_{113}$) 
correspond to the next-to-leading order term in the resummation of
$\Gamma^{(1)}$ in the first (second) line of Eq.~(\ref{bkexplicit}).
Lastly, his $B^{I}_{114}$ resums into $\Gamma^{(2)}$ of the first
line. 

\subsection{Multi-point propagators}
\label{sec:multipointpropagators}

In the previous section we showed how the $\Gamma^{(p)}$ 
functions in Eq.~(\ref{GpdefNG}) serve as
basic building blocks out of which one can construct series expressions for 
the polyspectra. In \cite{2008PhRvD..78j3521B} these functions were further identified as $n$-point propagators since
they can be obtained by functional differentiation of the final fields
with respect to the initial ones. For PNG the $\Gamma^{(p)}$ functions are
still equal to the ensemble average of the $p$-order functional derivative of $\Psi$ with respect to $\phi$, that is
\begin{equation}
\frac{1}{p!}\,
\mg\frac{\partial^p
\Psi_{i}(\vk,\eta)}{\partial\phi_{j_{1}}(\vq_{1})\dots\partial\phi_{j_{p}}(\vq_{p})}\md
\equiv
\Dirac(\vk-\vq_{1 \ldots p})\,\Gamma^{(p)}_{ij_{1}\dots
j_{p}}\left(\vq_{1},\dots,\vq_{p},\eta\right),
\label{GammaAllDef}
\end{equation}
where we have re-introduced the doublet indices for clarity. This can
be easily checked starting from Eq.~(\ref{mFndef}) and getting to Eq.~(\ref{GpdefNG}). Notice that,
contrary to the standard PT kernels in Eq.~(\ref{Fnkernels}), the multi-point
propagators depend on the statistics of the initial fields. 

\subsection{Multi-point propagators and correlation functions}

Within the framework developed so far, it is clear that a precise description of
the large scale clustering can be achieved provided with a good
understanding of the $\Gamma^{(p)}$ functions.
For Gaussian initial condition it has been observed that the
$\Gamma$-functions can be expressed in terms of
correlation functions between initial and final fields. For instance,
the nonlinear propagator $\Gamma^{(1)}$ satisfies \cite{2006PhRvD..73f3520C},
\begin{equation}
\mg\Psi_{i}(\vk,\eta)\,\phi_{j}(\vk')\md =\Dirac(\vk+\vk^\prime)\, \Gamma^{(1)}_{il}(k,\eta)\,P_{0,lj}(k),
\end{equation}
where $P_0$ is the power spectrum of the initial density field and
$\Gamma^{(1)}$ is defined formally through functional differentiation, as in
Eq.~(\ref{GammaAllDef}). An analogous expression is given in \cite{2008PhRvD..78j3521B} for $\Gamma^{(2)}$.
These relations played a key role for the Gaussian case since they allowed the measurement
of the fully nonlinear $\Gamma^{(1)}$ and $\Gamma^{(2)}$ function in N-body simulations \cite{2006PhRvD..73f3520C,2008PhRvD..78j3521B}.

These simple expressions however, are no longer valid for non-Gaussian initial conditions. In this case the cross-correlation function between $\Psi$ and $\phi$ can expanded as follows,
\begin{equation}
\mg\Psi(\vk,\eta)\,\phi(\vk')\md
= \sum_{p}\int\dd^3\vq_{1}\dots\dd^3\vq_{p}\,\Dirac(\vk-\vq_{1\dots p})\,\Gamma^{(p)}(\vq_{1},\dots,\vq_{p},\eta)
\langle \phi(\vq_{1})\dots\phi(\vq_{p})\phi(\vq')\rangle_{c}.
\label{psiphi}
\end{equation}
This is actually a peculiar case of the general expression as obtained in Eq. (\ref{PkExpansionNG}) where the sum is now restricted to $p_{2}=1$. This expression can then be written in terms of the spectra $P_0(k)$, bispectra $B_0(\vk_{1},\vk_{2})$ and
so on, of the initial fields times the corresponding $\Gamma^{(p)}$ function from Eq.~(\ref{GpdefNG}). It leads to,
\begin{eqnarray}
\mg\Psi_{i}(\vk,\eta)\,\phi_{j}(\vk')\md=\Dirac(\vk\!-\!\vk')\Big[&&\!\!\!\!\!\!
\Gamma^{(1)}_{il}(k,\eta)\,P_{0,lj}(k)+\nonumber\\
&&\!\!\!\int\dd^3\vq_{1}\dd^3\vq_{2}\,\Dirac(\vk\!-\!\vq_{12})\,
\Gamma^{(2)}_{ilm}(\vq_{1},\vq_{2},\eta)\,B_{0,lmj}(\vq_{1},\vq_{2})+\dots\Big]
\label{gamma1}
\end{eqnarray}

The terms appearing in the previous
equation are ordered in terms of importance for weakly non-Gaussian
models such as the local \cite{1990PhRvD..42.3936S,1994ApJ...430..447G,2000MNRAS.313..141V,2001PhRvD..63f3002K} and equilateral \cite{2004PhRvD..70h3005B} ones. Typically, the
initial curvature bispectrum in these models satisfies the
hierarchical\footnote{The extension of this scaling to the density spectra is however not straightforward because of the shape of the transfer function that relates the density and the potential at linear order.}  scaling $B_{\Phi}(k,k,k) \sim f_{NL} \, P^2_{\Phi}(k)$,
where $P_{\phi}(k)$ is the curvature power spectrum. The second term in Eq.~(\ref{gamma1}) is
thus sub-dominant by a factor $\sim f_{NL} D(z)$. Alternatively,
notice that Eq.~(\ref{gamma1}) resembles the expansion for $P(k,\eta)$ in Eq.~(\ref{pkexplicit}). This 
expansion have been studied using standard perturbation theory to
one-loop and the leading order induced by primordial non Gaussianity (i.e. the
second term in the previous equation) introduces
corrections of $\sim 1\%-2\%$ for both local and equilateral models
with $|f_{NL}| < 300$ \cite{2008PhRvD..78l3534T}.

Similarly the computation of $\mg\Psi_{i}(\vk,\eta)\,\phi_{j}(\vk_{1})\,\phi_{l}(\vk_{2})\md$ will make intervene the whole range of initial spectra. Its leading contribution is
\begin{eqnarray}
\mg\Psi_{i}(\vk,\eta)\,\phi_{j}(\vk_{1})\,\phi_{l}(\vk_{2})\md=\Dirac(\vk\!-\!\vk_{12})\Big[&&\!\!\!\!\!
\Gamma^{(1)}_{im}(k,\eta)\,B_{0,mjl}(\vk_{1},\vk_{2})\nonumber\\
&& + \ 2 \ \Gamma^{(2)}_{imn}(\vk_{1},\vk_{2},\eta)\,P_{0,mj}(k_{1})\,P_{0,nl}(0,k_{2})
+\dots\Big]\,.
\label{gamma2}
\end{eqnarray}
The relative importance of the terms in this series can be studied for
particular models but in general, for weakly non-Gaussian initial
conditions, the first two terms are dominant over the remaining series
(not shown). Again, this expansion resembles the $\Gamma$-expansion
for $B(k,k_1,k_2)$ in Eq.~(\ref{bkexplicit}). In turn, the bispectrum has been studied
using one-loop perturbation theory by \cite{2009PhRvD..80l3002S} for
local and equilateral models, and additionally using N-body measurements in \cite{2010MNRAS.tmp..721S}
for local models with $f_{NL}=\pm 100$. In these cases, the
next-to-leading terms become increasingly important for $k \simgt
0.1\,\Mpc$ depending on the value of $f_{NL}$ and redshift.

Therefore, if we neglect sub-leading terms it is still possible to
estimate the $\Gamma$-functions from correlators. This is well
justified for $\Gamma^{(1)}$, obtained from Eq.~(\ref{gamma1}) as,
\beq
\Dirac(\vk-\vk_1)\,\Gamma^{(1)}(k,\eta)=\mg\Psi(\vk,\eta)\,\phi(\vk_1)\md /P_{0}(k)\,.
\eeq
Inserting this into Eq.~(\ref{gamma2}) one recovers $\Gamma^{(2)}$ as,
\beq
\Dirac(\vk-\vk_{12})\Gamma^{(2)}(\vk_{1},\vk_{2},\eta) \simeq
\frac{\mg\Psi(\vk,\eta)\,\phi(\vk_{1})\,\phi(\vk_{2})\md}{2\,P_0(k_1)\,P_0(k_2)} - \frac{B_0(\vk_1,\vk_2)
\mg\Psi(\vk,\eta)\,\phi(\vk_1+\vk_2)\md }{2 \, P_{0}(k)\, P_{0}(k_1)\, P_{0}(k_2)}.
\label{gamma2bis}
\eeq
As expected, the first term is the same as for Gaussian initial statistics.
One can now insert this back into Eq.~(\ref{gamma1}) and recover the
correction introduced by the initial higher-order correlators such as 
$B_0$,
\beq
\Dirac(\vk-\vk_1)\,\Gamma^{(1)}(k,\eta)=\frac{\mg\Psi(\vk,\eta)\,\phi(\vk_1)\md}{P_{0}(k)}+\frac{1}{2P_{0}(k)}\int
d^3\vk_1 \,d^3\vk_2 \frac{\mg\Psi(\vk,\eta)\,\phi(\vk_{1})\,\phi(\vk_{2})\md}{ P_{0}(k_1)\, P_{0}(k_2)}\,B_0(\vk_1,\vk_2)+\mO(B_0^2).
\label{gamma1bis}
\eeq
In most commonly studied models of primordial non Gaussianity
\cite{1990PhRvD..42.3936S,2004PhRvD..70h3005B}, $B_0\sim f_{NL}$. The expression above is thus neglecting terms
of order $f_{NL}^2$. In addition, as we discuss in the next section, the cross-correlations
such as that in Eq.~(\ref{gamma1bis}) will drop to zero towards high
momenta. This is an important property to take into account when it comes to compute the momentum integration that cannot be recovered using the known tree-level analytical expression for these correlators.

Expressions such as Eq.~(\ref{gamma2bis}) and (\ref{gamma1bis}) give a concrete path to the
actual measurements of the $\Gamma$ functions that can certainly shed light
into their full description. This is however beyond the scope of this
paper and will be left for further work.

\section{Propagators in the large-$k$ limit}
\label{sec:largeklimit}

\subsection{The $\alpha$-method}

The multi-point propagators or $\Gamma$ functions play an essential role in
our formalism (as they did in \cite{2006PhRvD..73f3520C,2008PhRvD..78j3521B} for Gaussian initial conditions) and
setting the ground for their adequate analytical description is
thus one of the main goals of this paper. This task is however
extremely difficult as it involves summing over infinite terms in Eq.~(\ref{GpdefNG}) or adding up an infinite
number of diagrams, if we were to work with diagrammatic techniques as
in \cite{2006PhRvD..73f3519C,2008PhRvD..78j3521B} (but see also \cite{2007astro.ph..3563M,2008JCAP...10..036P}). Nonetheless well established physical arguments were
put forward in \cite{2006PhRvD..73f3520C} to show that indeed, the leading order set of
contributions can be resummed in the large-$k$ regime where the
coupling of Fourier modes simplifies (see below). The low-$k$ regime,
that allows perturbative calculations, could then be matched to the
high-$k$ asymptotic using simple physical arguments \cite{2006PhRvD..73f3520C}. This
resulted in a good agreement with propagator measurements in N-body
simulations and allowed accurate descriptions of power spectrum \cite{2008PhRvD..77b3533C} and bispectrum \cite{2008PhRvD..78j3521B} at large scales.

In what follows we are interested in extending these results to the case of non-Gaussian initial conditions, computing of the $\Gamma$ functions (defined as the
ensemble averages of functional derivatives of $\psi_{i}$ with respect to the initial field values $\phi_{j}$) in the large-$k$ limit. As shown in \cite{2006PhRvD..73f3520C,2008PhRvD..78j3521B} this is a regime where the other intervening modes $\vq$ in the
momentum (or ``loop'') integrals are such that $q\ll k_{i}$ for all ``external'' $\vk_{i}$. The intervening modes are also assumed to be in the linear regime with initial condition set in the growing mode. This later condition amounts to resum only those
diagrams that are maximally connected with the initial conditions
(that is, with all mode-mode interactions happening along the ``principal
path'' as introduced in \cite{2006PhRvD..73f3520C,2008PhRvD..78j3521B}) what gives
the leading contribution after resummation.

Following these assumptions we then compute ${\partial\Psi_{i}(\vk,\eta)}/{\partial\phi_{j}(\vk')}$ from Eq.~(\ref{eomi}) with the prescription that in its r.h.s. one replaces
one factor $\Psi_{k}$ by its linear solution
$g_{kl}(\eta',0)\phi_{l}$. Furthermore $\phi_{l}$ is
set is the growing mode, that is equal to
$\delta_{0}(\vq)\,u_l$ with $u_l=(1,1)$. From Eq.~(\ref{eomi}), one then has
\begin{equation}
 \frac{\partial \Psi_{i}(\vk,\eta)}{\partial\phi_{j}(\vk')}=
 g_{ij}(\eta,0)\,\Dirac(\vk-\vk')
 + 2\int_0^{\eta} \!\! {\dd \eta'}D_{+}(\eta')\, g_{ik}(\eta,\eta')\, \gamma_{klm}^{(\rm s)}(\vk,\vq,\vk_1)\,  u_{l}\,\delta_0(\vq)\, \frac{\partial \Psi_m(\vk_{1},\eta')}{\partial\phi_{j}(\vk')}\,,
\end{equation}
where the vertex function should be further taken in the $q\ll k_{1}$
limit (and recall that repeated Fourier arguments are assumed to be integrated over). As noticed in \cite{2006PhRvD..73f3520C} this considerably simplifies its expression and we are left with
\begin{equation}
 \frac{\partial \Psi_{i}(\vk,\eta)}{\partial\phi_{j}(\vk')}
 = g_{ij}(\eta,0)\,\Dirac(\vk-\vk')+\ii\,\alpha(\vk)\!\int_0^{\eta}\!\!  \dd\eta'{D_{+}(\eta')} \ g_{ik}(\eta,\eta')\,
 \frac{\partial \Psi_k(\vk,\eta')}{\partial\phi_{j}(\vk')}\,,
\end{equation}
where the intervening modes were all collected in a \textsl{single} quantity,
\begin{equation}
\alpha(\vk)=-\ii \int\dd^3\vq\,\frac{\vk.\vq}{q^2}\,\delta_{0}(\vq)\,,
\label{alpha}
\end{equation}
where $\delta_{0}(\vq)$ is the initial density contrast.

This possibility of collecting all the intervening modes in a single random variable (or more generally a finite number of variables) is at the heart of what we call here the $\alpha$-method. This method was described and extensively used in \cite{2008PhRvD..78h3503B} (and was already sketched in \cite{2007A&A...476...31V}). This method is much more powerful than standard diagram evaluations and countings. It eventually allows to compute the resulting propagator shapes either when the initial statistics is non trivial as in the present work and/or when the dependence of the propagators is complicated (as in \cite{2008PhRvD..78h3503B}).

Let us go back to this specific calculation. We note that for a given value of $\vk$, $\alpha(\vk)$ is a real random
variable. It is Gaussian distributed if the initial conditions are
Gaussian distributed but we will not make this hypothesis here. 
In the previous system we have introduced the $\ii$ factor to make $\alpha(\vk)$ real and of positive variance.  Note that in all cases,
\begin{equation}
\langle\alpha^2(\vk)\rangle=\int\dd^3\vq\ P_{0}(q)\,\frac{(\vq.\vk)^2}{q^4}=k^2\sigma_{\displ}^2,
\end{equation}
where $\sigma_{\displ}$ is the one-component \textsl{displacement}
dispersion in linear theory.

We are then left with an equation evolution for ${\partial \Psi_{i}(\vk,a)}/{\partial\phi_{j}(\eta')}$ that can be solved explicitly. If we indeed write 
\begin{equation}
 \frac{\partial \Psi_{i}(\vk,\eta)}{\partial\phi_{j}(\vk')}
 = \xi(\vk,\eta)\,g_{ij}(\tv,0)\,\Dirac(\vk-\vk')\,,
\end{equation}
taking advantage of Eq. (\ref{gcomp}), we are left with the ODE
\begin{equation}
\frac{\partial }{\partial \eta}\xi(\vk,\eta)
= \ii\,\alpha(\vk)\,D_{+}(\tv)\,\xi(\vk,\tv),\ \ \xi(\vk,\eta=0)=1\,,
\end{equation}
which can be solved explicitly to give, 
\begin{equation}
\frac{\partial \Psi_{i}(\vk,\eta)}{\partial\phi_{j}(\vk')}=\exp[\ii \alpha(\vk)(D_{+}(\eta)-1)]g_{ij}(\eta,0)\Dirac(\vk-\vk')\,.
\label{G1exp}
\end{equation}
This is the first result of this section. We will see later its implication for the form of the propagators.

We can then proceed to the computation of higher order $\Gamma$ functions from successive differentiations of Eq.~(\ref{eomi}) with respect to $\phi$. The second order partial derivative induces two terms. 
\begin{eqnarray}
 \frac{\partial^2 \Psi_{i}(\vk,\eta)}{\partial\phi_{j}(\vk_{1})\,\partial\phi_{k}(\vk_{2})}
 &=&
\!\int_0^{\eta}\!\!  {\dd\eta'} \ g_{il}(\eta,\eta')\, \gamma_{lmn}^{(\rm s)}(\vk,\vk'_{1},\vk'_2)\,
\frac{\partial \Psi_m(\vk'_{1},\eta')}{\partial\phi_{j}(\vk_{1})}\,
\frac{\partial \Psi_n(\vk'_{2},\eta')}{\partial\phi_{k}(\vk_{2})}\nonumber\\
&&+\, 2\int_0^{\eta}  \!\! \dd\eta'{D_{+}(\eta')} \ g_{il}(\eta,\eta')\, \gamma_{lmn}^{(\rm s)}(\vk,\vq,\vk')\,  \phi_{m}(\vq)\,\frac{\partial^2 \Psi_n(\vk',\eta')}{\partial\phi_{j}(\vk_{1})\,\partial\phi_{k}(\vk_{2})},
\end{eqnarray}
which can once again be simplified in the high-$k$ limit into,
\begin{eqnarray}
 \frac{\partial^2 \Psi_{i}(\vk,\eta)}{\partial\phi_{j}(\vk_{1})\partial\phi_{k}(\vk_{2})}&=&
\int_0^{\eta}   {\dd \eta'} \ g_{il}(\eta,\eta') \gamma_{lmn}^{(\rm s)}(\vk,\vk'_{1},\vk'_2)
\frac{\partial \Psi_m(\vk'_{1},\eta')}{\partial\phi_{j}(\vk_{1})}
\frac{\partial \Psi_n(\vk'_{2},\eta')}{\partial\phi_{k}(\vk_{2})}\nonumber\\
&&+ \ii\alpha(\vk)\int_0^{\eta} \dd\eta'{D_{+}(\eta')} \ g_{il}(\eta,\eta') \frac{\partial^2 \Psi_l(\vk',\eta')}{\partial\phi_{j}(\vk_{1})\partial\phi_{k}(\vk_{2})}.
\label{eomG22}
\end{eqnarray}

It can easily be checked that the solution of Eq.~(\ref{eomG22}) is 
\begin{equation}
\frac{\partial^2 \Psi_{i}(\vk,\eta)}{\partial\phi_{j}(\vk_{1})\,\partial\phi_{k}(\vk_{2})}=
\exp[\ii \alpha(\vk)(D_{+}(\eta)-1)]\int_0^{\eta}   {\dd \eta'}\, \ g_{il}(\eta,\eta') \,\gamma_{lmn}^{(\rm s)}(\vk,\vk_{1},\vk_2)\,g_{mj}(\eta',0)\,g_{nk}(\eta',0),
\label{G2Exp}
\end{equation}
where the integral in the left hand part of this equation is the expression of 
$\partial^2 \Psi_{i}(\vk,a)\,/\,\partial\phi_{j}(\vk_{1})\partial\phi_{k}(\vk_{2})$ 
for a vanishing value of $\alpha$ that is  its tree order expression. 
Indeed, inserting this expression in the second term of the right hand side of Eq.~(\ref{eomG22}) and using the result of Eq.~(\ref{G1exp}), we are left with
\begin{eqnarray}
 \frac{\partial^2 \Psi_{i}(\vk,\eta)}{\partial\phi_{j}(\vk_{1})\,\partial\phi_{k}(\vk_{2})}&=&
\!\int_0^{\eta}\!\! {\dd \eta'}\ g_{il}(\eta,\eta') \gamma_{lmn}^{(\rm s)}(\vk,\vk_{1},\vk_2)
\exp[\ii \alpha(\vk_{1}+\vk_{2})(D_{+}(\eta')-1)]g_{mj}(\eta',0)\,
g_{nk}(\eta',0)+
\nonumber\\
&&\hspace{-3cm}\ii\alpha(\vk)\!\int_{\eta''}^{\eta}\!\! \dd\eta'{D_{+}(\eta')} \ g_{il}(\eta,\eta')\exp[\ii \alpha(\vk)(D_{+}(\eta')\!-\!1)]
\!
\int_0^{\eta}\!\!  {\dd \eta''}\ g_{lq}(\eta',\eta'')\, \gamma_{qmn}^{(\rm s)}(\vk,\vk_{1},\vk_2)\,g_{mj}(\eta'',0)\,g_{nk}(\eta'',0).
\label{eomG22b}
\end{eqnarray}
Then remarking that 
\begin{eqnarray}
\ii\,\alpha(\vk)\!\int_{\eta''}^{\eta}\!\! {\dd \eta'}D_{+}(\eta')\, g_{il}(\eta,\eta')\exp[\ii \alpha(\vk)(D_{+}(\eta')-1)]\,g_{lq}(\eta',\eta'') =\nonumber\\
g_{iq}(\eta,\eta'')
\Big\{\exp[\ii\alpha(\vk)(D_{+}(\eta)\!-\!1)]-\exp[\ii\alpha(\vk)(D_{+}(\eta')\!-\!1)]\Big\},\label{xirelation}
\end{eqnarray}
and that $\alpha(\vk)$ is a linear function of its argument,
the two terms of the right hand side of Eq.~(\ref{eomG22b}) simplify and recombine into the expression of Eq.~(\ref{G2Exp}).

This result can naturally be extended to any order\footnote{An alternative way of obtaining this result is to remark that in presence of low $q$ perturbing modes Eq.~(\ref{eom}) can be rewritten -- for an Einstein-de Sitter background -- as
$a\frac{\partial}{\partial a} \Psi_i(\vk,a) + \Omega_{ij} \Psi_j(\vk,a)-\ii\alpha(\vk) a\Psi_i(\vk,a)=
\gamma_{ijk}(\vk,\vk_1,\vk_2) \ \Psi_j(\vk_1,a) \ \Psi_k(\vk_2,a).$
The linear progator is
$\tilde{g}_{ij}(\vk,a_{1},a_{2})=\exp(\ii\alpha(\vk)(a_{1}-a_{2}))g_{ij}(a_{1}/a_{2})$. The quantities we derive in this paper are the tree order $\Gamma$-functions of this theory, whose $\alpha$ dependence can be easily computed from the fact that
$\tilde{g}_{ij}(\vk_{1}+\vk_{2},a_{1},a_{2})\tilde{g}_{kl}(\vk_{1},a_{2},a_{3})\tilde{g}_{mn}(\vk_{2},a_{2},a_{3})=
\exp(\ii\alpha(\vk_{1}+\vk_{2})(a_{1}-a_{3})){g}_{ij}(a_{1}/a_{2}){g}_{kl}(a_{2}/a_{3}){g}_{mn}(a_{2}/a_{3})$.}
as it can be established recursively by successive use of the relation (\ref{xirelation}). More precisely, we have
\begin{equation}
\frac{\partial^p \Psi_{i}(\vk,\eta)}{\partial\phi_{j_{1}}(\vk_{1})\dots\partial\phi_{j_{p}}(\vk_{p})}=
\exp\left[\ii \alpha\left(\vk\right)(D_{+}-1)\right]\left.\frac{\partial^p \Psi_{i}(\vk,\eta)}{\partial\phi_{j_{1}}(\vk_{1})\dots\partial\phi_{j_{p}}(\vk_{p})}\right\vert_{\alpha(\vk)=0},
\label{GpExp}
\end{equation}
where $\vk = \sum\vk_{j}$. After taking the ensemble average of this
expression we finally obtain,
\beq
\Gamma^{(p)}(\vk_1,\dots,\vk_p,\eta) = \langle\,\exp\left[\ii
  \alpha\left(\vk\right)(D_{+}(\eta)-1)\right]\,\rangle\,\Gamma^{(p)}_{\rm tree}(\vk_1,\dots,\vk_p,\eta)\,,
\eeq
where we have used that the derivatives on the r.h.s of Eq.~(\ref{GpExp}) are
evaluated at $\alpha=0$ (i.e. at tree level) and thus are
independent of initial random fields. This is a remarkable result that
extends in a concrete and simple way the findings of \cite{2006PhRvD..73f3520C,2008PhRvD..78j3521B} for Gaussian initial conditions.
Hence, the statistical properties of ``random variable'' $\alpha(\vk)$
determine a transition function that relates the fully nonlinear
multi-point propagator with their tree level expression (which, up to
sub-leading terms, are nothing else than the standard PT kernels in Eq.~\ref{Fnkernels}), in the
high-$k$ regime. In what follows, we will use this result to explicitly
compute the $\Gamma$-functions in this regime.

\subsection{The transition function}

Irrespectively of their order, the $\Gamma$-functions are all obtained, in the large-$k$ regime, from the computation of the ensemble average in
\begin{equation}
f(k)=\langle\,\exp\left[\ii\alpha(\vk)(D_{+}-1)\right]\,\rangle\,,
\end{equation}
that will depend only on $k$ due to homogeneity and isotropy. 
From the definition of $\alpha(\vk)$ in Eq.~(\ref{alpha})
we observe that the function $f(k)$ can in fact be easily related to
the statistical properties of the primordial field: it is nothing but the moment generating function $\mM$ (see Eq.~(133) of \cite{2002PhR...367....1B}) at point $t=\vk(D_{+}-1)$ of the single point one component displacement field $\vd$,
\begin{equation}
\vd=\int\dd^3\vq\,\frac{\vq}{q^2}\,\delta_{0}(\vq).
\label{disp}
\end{equation}
As a consequence of Eq.~(135) in \cite{2002PhR...367....1B}, $\log
f(k)$ is nothing but the cumulant generating function of that same
variable at the same point defined through,
\begin{equation}
\log
f(k)=\sum_{p=2}^{\infty}\frac{\langle(\vd\cdot\vk)^p\rangle_{c}}{p!}(D_{+}-1)^p.
\label{fk}
\end{equation}
For Gaussian initial conditions only $p=2$ is non-zero and one recovers the well known result $f(k)=\exp(-k^2\sigma_{\displ}^2(D_{+}-1)^2/2 )$
\cite{2006PhRvD..73f3520C,2008PhRvD..78j3521B}. 
Note also that only even values of $p$ contribute to this sum because the cumulants cannot depend on the direction of the wave-vector $\vk$.
This is at variance with
the result derived in \cite{2007PhRvD..76h3517I} following the
prescription for propagator resummation put forward in
\cite{2007astro.ph..3563M} for Gaussian initial statistics.
It is however a very important result because it shows that, for any
given model of primordial non-Gaussianity, the first ``non-Gaussian''
correction to the propagator decay (in the high-$k$ limit) is given by 
the four point connected function. For weakly non-Gaussian initial conditions this will
represent a minor contribution.

\section{Predictions for the local model of PNG}
\label{sec:localmodel}

We now turn into the evaluation of the transition function $f(k)$
defined in Eq.~(\ref{fk}) for a specific model of primordial non-Gaussianity. We will focus in the local model \cite{1990PhRvD..42.3936S,1994ApJ...430..447G,2000MNRAS.313..141V,2001PhRvD..63f3002K},
which is perhaps the most studied model of primordial non Gaussianity within the
context of large scale structure (see for instance the recent reviews \cite{2010arXiv1001.4707L,2010arXiv1001.5217V} and
references therein). This model is build upon the nonlinear 
relation \cite{1990PhRvD..42.3936S}
\beq
\Phi(\vx)=\phi(\vx)+f_{NL}(\phi^2(\xv)-\langle \phi^2(\vx)
\rangle)+g_{NL} \phi^3(\vx)\,,
\eeq
between the Gaussian field $\phi$ and Bardeen's gauge-invariant potential $\Phi$, where up to cubic terms are considered. The local model corresponds to $f_{NL}$ and $g_{NL}$ constants independent of space. The connection to the matter overdensity
is through Poisson's equation $\delta_{\vk}(z)=M(k,z)\Phi_{\vk}$, with
\beq
M(k,z)=\frac{2}{3}\frac{k^2T(k)D(z)}{\Omega_mH_0^2}\,,
\eeq
where $T(k)$ is the matter transfer function, and $D(z)$ the growth
factor. In what follows we will assume the same cosmological model as
in \cite{2010MNRAS.tmp..721S} (and the CMB convention for $f_{NL}$ and
$g_{NL}$). This can be summarized as $h=0.7$, $\Omega_m=0.279$,
$\Omega_b=0.0462$, $n_s=0.96$ and $\sigma_8\simeq 0.81$ (see Sec.~3
in \cite{2010MNRAS.tmp..721S} for more detail)

We start by writing the first terms of the sum defining $f(k)$,
\beq
\log f(k) = \frac{\la(\vd\cdot\kv)^2\ra_c}{2}(D_+-1)^2+\frac{\la(\vd\cdot\kv)^4\ra_c}{24}(D_+-1)^4+\dots,
\label{logf}
\eeq
with the displacement field $\vd$ given by Eq.~(\ref{disp}).
As noticed before, the first contribution reduces to
$\la(\vd\cdot\kv)^2\ra_c=-\sigma_{\rm disp}^2k^2$ and
$\sigma_{\rm disp} = 6.01 \Mpc$ (at $z=0$) for our cosmology. For 
Gaussian initial conditions this is the only non-zero contribution.
For non-Gaussian initial conditions we need to evaluate, in addition
\bea
\la(\vd\cdot\kv)^4\ra_c
& = & \int\!\! \de^3q_1\!\int\!\! \de^3q_2\!\int\!\! \de^3q_3\!\int\!\! \de^3q_4\frac{\kv\cdot\qv_1}{q_1^2}\,\frac{\kv\cdot\qv_2}{q_2^2}\,\frac{\kv\cdot\qv_3}{q_3^2}\,\frac{\kv\cdot\qv_4}{q_4^2}\,\la\d_0(\qv_1)\d_0(\qv_2)\d_0(\qv_3)\d_0(\qv_4)\ra_c\nonumber\\
& = & \int\!\! \de^3q_1\!\int\!\! \de^3q_2\!\int\!\! \de^3q_3\frac{\kv\cdot\qv_1}{q_1^2}\,\frac{\kv\cdot\qv_2}{q_2^2}\,\frac{\kv\cdot\qv_3}{q_3^2}\,\frac{\kv\cdot\qv_4}{q_4^2}
~T_0(\qv_1,\qv_2,\qv_3,\qv_4)\,,
\label{dk4}
\eea
with $\qv_4=-\qv_{123}\equiv-(\qv_1+\qv_2+\qv_3)$ in the last equality. The matter trispectrum $T_0$ is given by
\bea
T_0(\kv_1,\kv_2,\kv_3,\kv_4) & = & M(k_1,z)M(k_2,z)M(k_3,z)M(k_4,z)~ T_\Phi(\kv_1,\kv_2,\kv_3,\kv_4)\,,
\label{eq:T0}
\eea
where $T_\Phi$ is the curvature trispectrum.
For local non-Gaussianity there are two distinct contributions,
\bea
\label{eq:Tlc}
T_\Phi(\kv_1,\kv_2,\kv_3,\kv_4) & = & 4 \fNL^2 P_\Phi(k_1)P_\Phi(k_2) \left[P_\Phi(k_{13})+P_\Phi(k_{14})\right]+ {\rm 5~perm.}\nonumber\\
& & +6 \gNL P_\Phi(k_1)P_\Phi(k_2)P_\Phi(k_{3})+ {\rm 3~perm.}\,.
\eea
Hence, in what follows we will distinguish the $\fNL^2$ component from
the $\gNL$ one. In appendix \ref{integraleval} we include a detailed
account of the evaluation of the integral in Eq.~(\ref{dk4}). The
final result for our cosmology is $\la(\vd\cdot\kv)^4\ra_c= f^2_{NL} k^4 s^4_{4,f_{NL}}+ g_{NL} k^4 s^4_{4,g_{NL}}$
with $s_{4,f_{NL}}=0.05537\Mpc$ and $s_{4,g_{NL}}=0.02359\Mpc$
(at $z=0$). After putting together this result and the one for the variance
 into Eq.~(\ref{logf}), and normalizing the growth to $D(z=0)=1$, we
obtain the prediction for the transition function in the local model
at arbitrary redshift. 

\begin{figure}[t]
\begin{center}
{\includegraphics[width=0.45\textwidth]{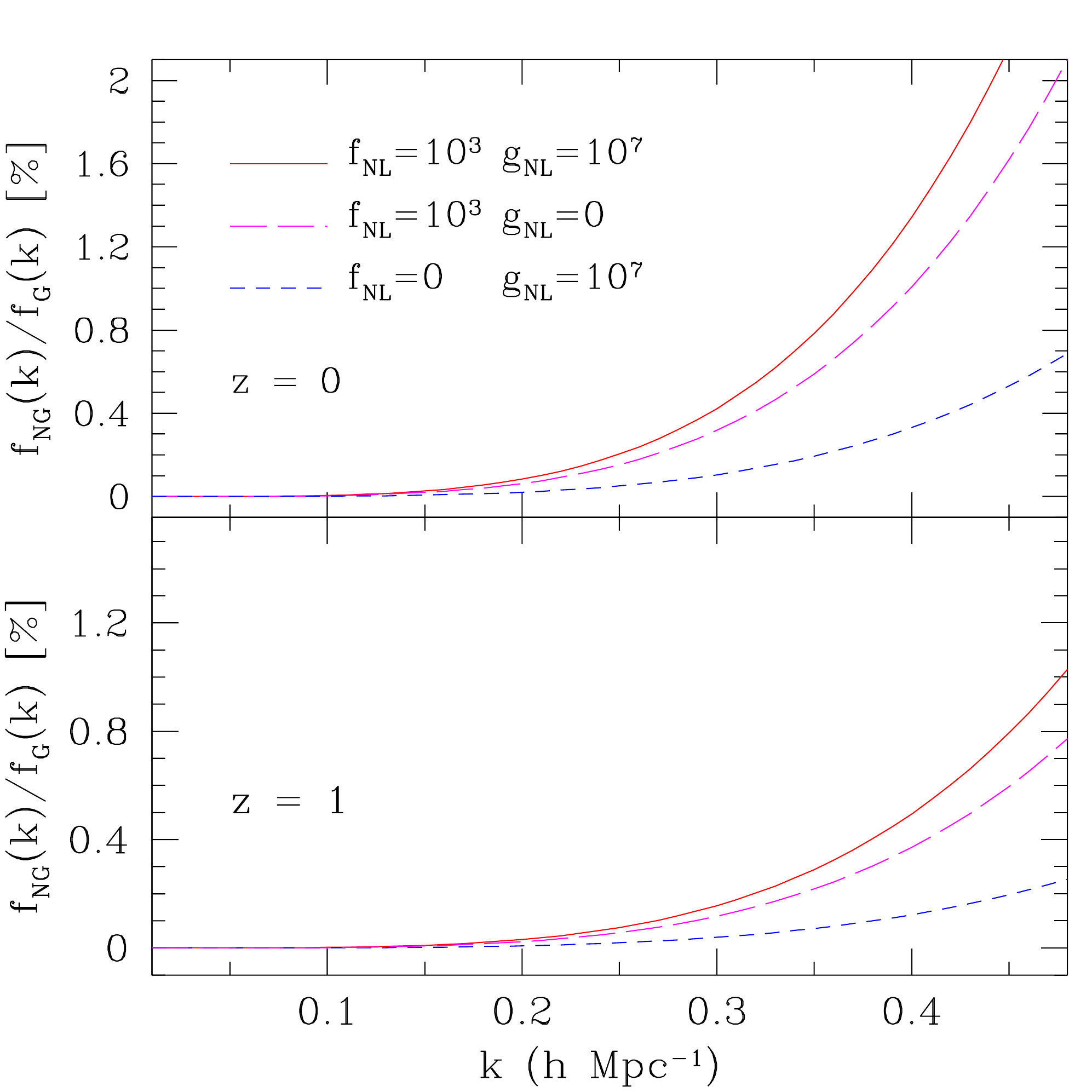}}
\caption{The non Gaussian transition function $f(k)$ in the large-$k$
  limit in ratio to its expression for Gaussian statistics
  (i.e. setting $\fNL=\gNL=0$). For currently accepted values of non Gaussian
  parameters (i.e. close to those shown) the correction to the damping
  is remarkably sub-dominant,
  even more so at higher redshift (e.g. bottom panel).}
\label{fig:fkratio}
\end{center}\end{figure}

\begin{figure}[t]
\begin{center}
{\includegraphics[width=0.45\textwidth]{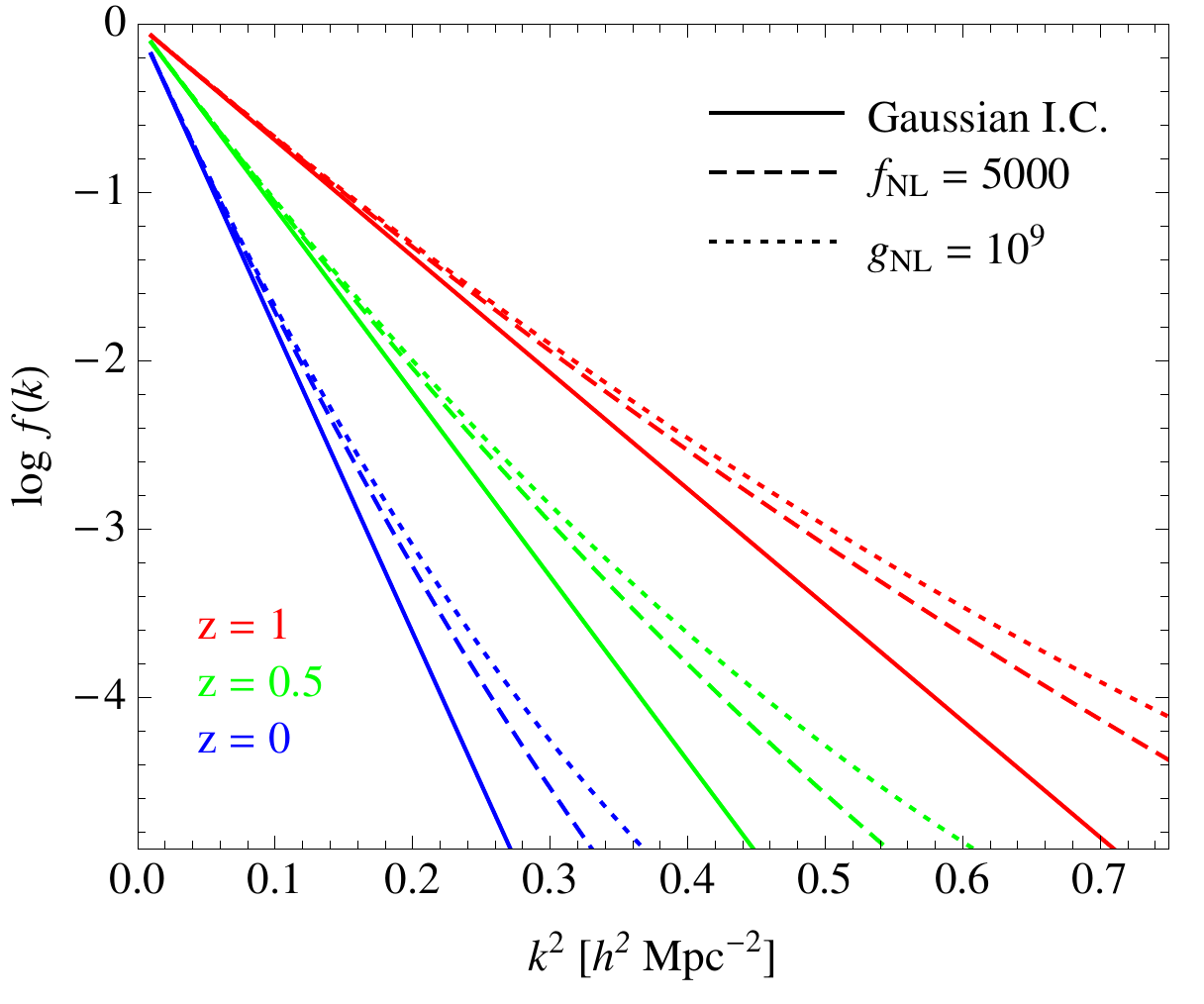}}
\caption{The damping function $\log f(k)$ as a function of $k$ for
Gaussian and non-Gaussian initial conditions, with $\fNL= 5\times 10^3$ and
$\gNL=10^9$ respectively}
\label{fig:a}
\end{center}\end{figure}

Figure~\ref{fig:fkratio} shows the ratio of the $f(k)$ function so
obtained, assuming $\fNL=10^3$ and/or $\gNL=10^7$, to its value for Gaussian initial conditions (i.e. with the kurtosis set to zero). We see that the modification to the Gaussian case for $f(k)$ is minor for this choice of the PNG parameters, with up to $2\%$ weaker decay at $z=0$ (top panel) and $1\%$ at $z=0.5$ (bottom panel).
In turn, Fig.~\ref{fig:a} shows $\log f(k)$ as a function of $k^2$ for
Gaussian and non-Gaussian initial conditions, with $\fNL= 5\times 10^3$ and, separately,
$\gNL=10^9$. As expected, the correction becomes relevant for values of $\fNL$ that make the $\phi^2$ correction the same order of the Gaussian component.

\begin{figure}[htbp]
\begin{center}
{\includegraphics[width=0.45\textwidth]{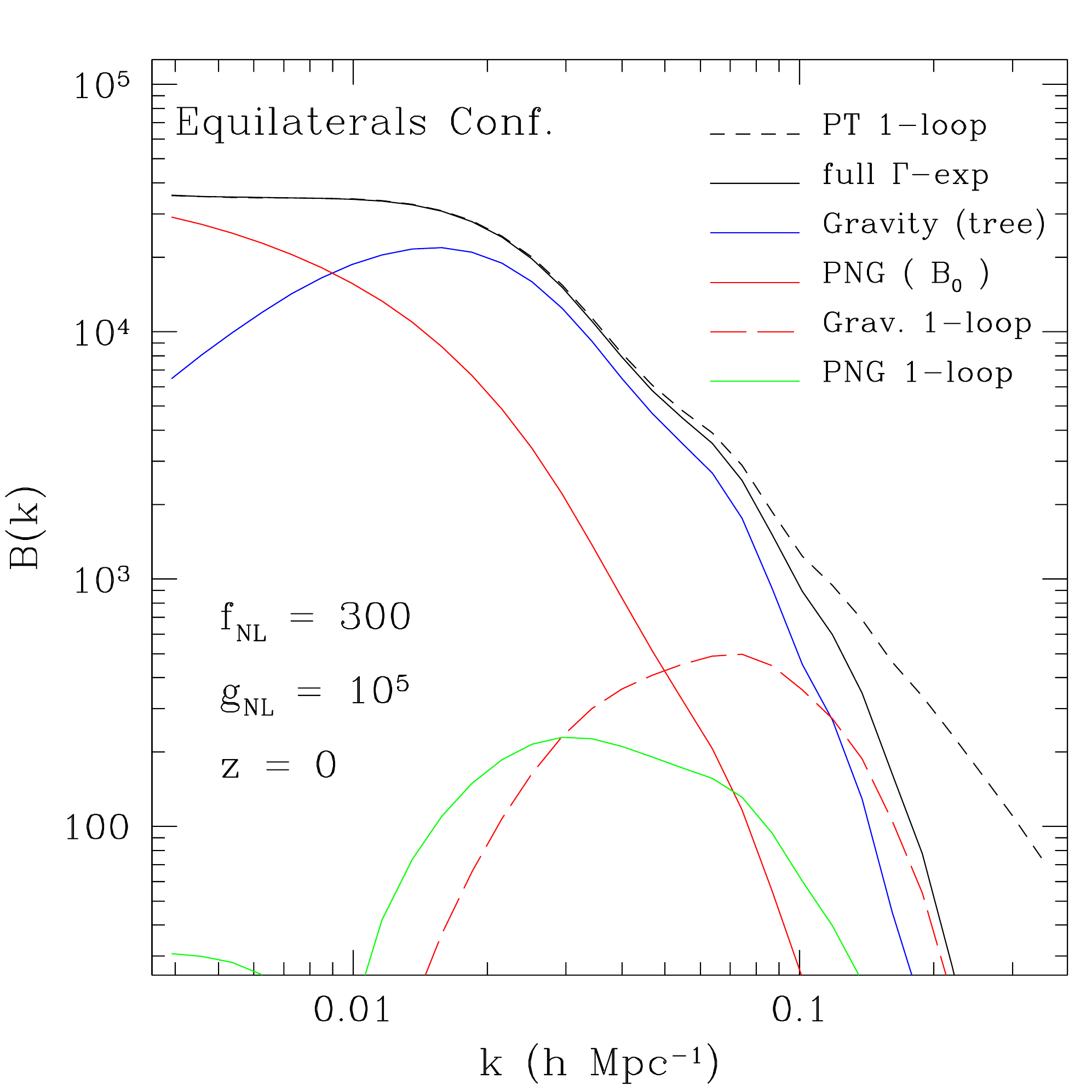}}
\caption{The $\Gamma$ expansion for the equilateral bispectrum in the local model
  of PNG, up to one-loop. The ``gravitational'' tree level bispectrum is shaped by
  $\Gamma^{(2)}$ while the PNG one arises from the primordial
  bispectrum and is exponentially suppressed to high-$k$ by $\Gamma^{(1)}$.
At one-loop there are terms $\sim P_0^3$ (grav.), $\sim P_0\times B_0$
(PNG, $f_{\rm NL}$) and $\sim T_0$ (PNG, $f^2_{\rm NL}$ and $g_{\rm NL}$).}
\label{fig:bisp}
\end{center}\end{figure}

Provided with the damping function $f(k)$ we are in place to
explicitly compute the $\Gamma$ -expansion for the
bispectrum, as given in Eq.~(\ref{bkexplicit}), but assuming the
multi-point propagators in their large-$k$ limit (a well justified approximation, at least for Gaussian initial conditions \cite{2008PhRvD..78j3521B}). Figure \ref{fig:bisp} shows the terms contributing to this expansion assuming a local model for primordial non-Gaussianity with  $\fNL=300$ and $\gNL=10^5$, and equilateral configurations. In solid blue line we show the equivalent to the tree-level bispectrum induced by gravity, i.e. the first term in Eq.~(\ref{bkexplicit}). The contribution from the primordial bispectrum is depicted by solid red line, it dominates the total signal at very large scales but it is exponentially suppressed at high-$k$ by $\Gamma^{(1)}$. At one-loop there are two terms also present for Gaussian initial conditions, see discussion after
 Eq.~(\ref{bkexplicit}), which are $\mathcal{O}(P_0^3)$. These are shown by the dashed red line. The purely PNG induced ones at this order are given by the solid green line, they compromise terms $\mathcal{O}(P_0 B_0) \sim f_{\rm NL} P_0^3$ and $\mathcal{O}(T_0) \sim (f^2_{\rm NL}, g_{\rm NL}) P_0^3$. In dashed line we show the result of standard perturbation theory at one-loop from \cite{2010MNRAS.tmp..721S}. Contrary to that case, here the number of terms is reduced (due to the resummation of several
 contributions). In addition, each term dominates over a narrow range
 of scales. 

\section{Conclusions}
\label{sec:conclusions}

The impact of primordial non-Gaussian (PNG) initial conditions on the
statistical properties of the cosmic density field is a priori not
easy to infer. This is because most of the diagrammatic expansions
that have been lately used to study the large scale clustering are
rooted in the assumption of primordial Gaussian fluctuations, that
considerably simplifies the structure of multi-loop diagrams.
We show here explicitly that PNG can indeed be accounted for in the
approaches developed in previous papers along the so-called RPT method
\cite{2006PhRvD..73f3519C}. In particular the $\Gamma$-expansion approach of \cite{2008PhRvD..78j3521B} is preserved for
such initial conditions. 

Within this context we generalize the definition of multi-point propagator to the case of
arbitrary initial statistics.  We find that they can still be regarded
as the basic building blocks out of which one can construct $n$-point
spectra. The $\Gamma$-expansions so obtained correspond to a resummation of
infinite sub-sets of diagrams in the approach of \cite{2006PhRvD..73f3519C,2008ApJ...674..617T,2007astro.ph..3563M}, concretely those 
corresponding to propagator and vertex renormalization, in line with the results in \cite{2008PhRvD..78j3521B} for initial Gaussian fields. For
concreteness we give explicit expressions for the series expansion up
to one-loop of the power spectrum and the bispectrum. 
In this way, the leading corrections to the power-spectrum are then those that make intervene the primordial bispectrum in the contraction of the $\Gamma^{(1)}-\Gamma^{(2)}$ product.

We then turned into the description of the $\Gamma^{(n)}$ functions
themselves (a.k.a. the multi-point propagators). We show how the
simple one-to-one correspondence between $n$-point propagators and
$n$-point correlators (among initial and final fields) found for
Gaussian initial conditions can be extended to the case of PNG. In the latter case, however, we find that this relations involves a infinite hierarchy among $\Gamma^{(n)}$ and correlators. Nonetheless for weakly non-Gaussian models, as those allowed by current data, well justified approximations can be put forward to close the hierarchy permitting the actual measurement of $\Gamma^{(n)}$ in N-body simulations. In this
way one can extend the methodology carried out for Gaussian initial statistics \cite{2006PhRvD..73f3520C,2008PhRvD..78j3521B} to shed light in the full description of the multi-point propagators.

Finally, the large-$k$ damping behavior of the multi-point propagators
is derived for arbitrary initial conditions. In
this regime, we find that the multi-point propagators are proportional to
their tree-level expressions. Moreover we explicitly show
that the rate of decay is the same irrespectively of their order, and
that is given by the cumulant generating function of the displacement
field. Remarkably, the first corrections to the ``Gaussian''
damping in this regime is due to the initial trispectrum, which for small departure from Gaussian initial conditions constitutes a sub-dominant contribution.

We leave for future work a more detailed analysis involving quantitative
predictions for specific models of PNG. But we hope the results
already presented here might serve as a basis to meet the accuracy requirements demanded by the analysis of future datasets tailored at deciphering the
primordial statistics and the physics of inflation.

\section*{Acknowledgments}
This work was supported in part by the French Programme National de
Cosmology and by the French Agence National de la Recherche under
grant BLAN07- 1-212615. 
MC acknowledges support from Spanish Ministerio de Ciencia y Tecnologia (MEC), project AYA2006-06341 and the Juan de la Cierva MEC program.
ES acknowledges support from the European Commission under the Marie Curie Inter European Fellowship.

\appendix

\section{Evaluation of the kurtosis cumulant}
\label{integraleval}

In this appendix we discuss the evaluation of the $4$-order cumulant
defining the first non-Gaussian contribution to the transition function $f(k)$ in Eq.~(\ref{fk}). As discussed in Sec.~\ref{sec:localmodel}, and particularly in Eqs.~(\ref{eq:T0},\ref{eq:Tlc}), we must evaluate two separate contributions
to the trispectrum defining the kurtosis cumulant. We have, in the first place
\beq
\la(\vd\cdot\kv)^4\ra_{c,\fNL}
 =  \int\!\! \de^3q_1\!\int\!\! \de^3q_2\!\int\!\! \de^3q_3\,\frac{\kv\cdot\qv_1}{q_1^2}\,\frac{\kv\cdot\qv_2}{q_2^2}\,\frac{\kv\cdot\qv_3}{q_3^2}\,\frac{\kv\cdot\qv_4}{q_4^2}
~\widetilde{T}_{\fNL}(\qv_1,\qv_2,\qv_3,\qv_4)\,,
\eeq
with $\vq_4=-\vq_{123}$ and
\beq
\widetilde{T}_{\fNL}(\qv_1,\qv_2,\qv_3,\qv_4)=48\,\fNL^2\,M(q_1)\,M(q_2)\,M(q_3)\,M(q_4)\, P_\Phi(q_1)\,P_\Phi(q_3)\,P_\Phi(q_{12})\,,
\label{Tfnl}
\eeq
giving the $\fNL$ term, and a similar expression with the trispectrum replaced by
\beq
\widetilde{T}_{\gNL}(\vq_1,\vq_2,\vq_3,\vq_4)=24\,\gNL\,M(q_1)\,M(q_2)\,M(q_3)\,M(q_4)
\,P_\Phi(q_1)\,P_\Phi(q_2)\,P_\Phi(q_3)\,,
\label{Tgnl}
\eeq
corresponding to the $\gNL$ contribution. Notice that in Eqs.~(\ref{Tfnl},\ref{Tgnl})
we took advantage of the fact that we are integrating over all
orientations so all permutations give the same result. 
Since both integrals depend only on the magnitude $k$, we can integrate over the
orientation of $\kv$. Hence, for the $\fNL$ contribution we have
\beq
\la(\vd\cdot\kv)^4\ra_{c,\fNL}
 = \! \int \!\!\de^3q_1\!\int\!\! \de^3q_2\!\int \!\!\de^3q_3
\,\widetilde{T}_{\fNL}(\qv_1,\qv_2,\qv_3,\qv_4)\frac{1}{4\pi}\!\int\!\! \de \Omega_k~\frac{\kv\cdot\qv_1}{q_1^2}\frac{\kv\cdot\qv_2}{q_2^2}\frac{\kv\cdot\qv_3}{q_3^2}\frac{\kv\cdot\qv_4}{q_4^2},
\eeq
and we are then allowed to put $\qv_1$ along the $z$-axis and set the azimuthal angle of $\qv_2$ equal to zero, obtaining the 6-dimensional integral
\bea
\la(\vd\cdot\kv)^4\ra_{c,\fNL}
 & = & 2\pi\!\!\int\!\! \de q_1 \,q_1^2\!\int\!\! \de q_2\, q_2^2\!\int\!\!\de\cos\theta_2\!\int \!\!\de^3q_3
\,\widetilde{T}_{\fNL}(\qv_1,\qv_2,\qv_3,\qv_4)\!\int\!\! \de \Omega_k~\frac{\kv\cdot\qv_1}{q_1^2}\frac{\kv\cdot\qv_2}{q_2^2}\frac{\kv\cdot\qv_3}{q_3^2}\frac{\kv\cdot\qv_4}{q_4^2},
\nonumber \\
 & = & 2\pi\!\!\int \!\!\de q_1 \!\!\int \de q_2\!\int\!\!\de\cos\theta_2\!\int\!\! \de^3q_3
~\frac{\widetilde{T}_{\fNL}(q_1,q_2,q_3,q_4,q_{12})}{q_3^2 q_4^2}\!\int\!\!
\de
\Omega_k~(\kv\cdot\qv_1)(\kv\cdot\qv_2)(\kv\cdot\qv_3)(\kv\cdot\qv_4). \nonumber
\eea
Introducing the vector $\pv\equiv\qv_{12}\equiv\qv_1+\qv_2$ we can rewrite the integral as
\bea
\la(\vd\cdot\kv)^4\ra_{c,\fNL}
 & = & 2\pi\!\!\int \de q_1 \!\!\int \de q_2 \!\!\int_{|q_1-q_2|}^{q_1+q_2}\de p\frac{p}{q_1q_2}\!\!\int \de^3q_3
~\frac{\widetilde{T}_{\fNL}(q_1,q_2,q_3,q_4,p)}{q_3^2 q_4^2}\int \de \Omega_k~(\kv\cdot\qv_1)(\kv\cdot\qv_2)(\kv\cdot\qv_3)(\kv\cdot\qv_4),
\nonumber
\eea
and recall that $\qv_4=-\qv_{123}=-\qv_{12}-\qv_3=-\pv-\qv_3$, so that the magnitude $q_4$ only depends on the magnitudes of $\pv$ and $\qv_3$ and on the angle between them. Then, we can transform the integrand in the basis where $\pv=\qv_{12}$ is along the $z$-axis, 
\bea
\la(\vd\cdot\kv)^4\ra_{c,\fNL}
 & = & 2\pi\!\!\int \de q_1 \!\!\int \de q_2 \!\!\int_{|q_1-q_2|}^{q_1+q_2}\de p\frac{p}{q_1q_2}\!\!\int \de q_3 q_3^2\!\!\int\de\cos\tilde{\theta}_3
~\frac{\widetilde{T}_{\fNL}(q_1,q_2,q_3,q_4,p)}{q_3^2 q_4^2}\nonumber\\
& &\times \int \de \tilde{\phi}_3\int \de \Omega_k~(\kv\cdot\qv_1)(\kv\cdot\qv_2)(\kv\cdot\qv_3)(\kv\cdot\qv_4),
\eea
with $\tilde{\theta}_3$ and $\tilde{\phi}_3$ define the orientation of $\qv_3$ in the new basis and $\tilde{\theta}_3$ is now the angle between $\qv_3$ and $\pv$. Given the one-to-one correspondence between $\tilde{\theta}_3$ and $q_4$, as before we can change variable of integration to get
\bea
\la(\vd\cdot\kv)^4\ra_{c,\fNL}
 & = & 2\pi\!\!\int \de q_1 \!\!\int \de q_2 \!\!\int_{|q_1-q_2|}^{q_1+q_2}\de p\frac{p}{q_1q_2}\!\!\int \de q_3 \!\!\int_{|p-q_3|}^{p+q_3}\de q_4\frac{q_4}{pq_3}
~\frac{\widetilde{T}_{\fNL}(q_1,q_2,q_3,q_4,p)}{q_4^2}\nonumber\\
& &\times \int \de \tilde{\phi}_3\int \de \Omega_k~(\kv\cdot\qv_1)(\kv\cdot\qv_2)(\kv\cdot\qv_3)(\kv\cdot\qv_4)\\
 & = & 2\pi k^4 \!\!\int \de q_1 \!\!\int \de q_2 \!\!\int_{|q_1-q_2|}^{q_1+q_2}\de p\!\!\int \de q_3 \!\!\int_{|p-q_3|}^{p+q_3}\de q_4
 ~\frac{\widetilde{T}_{\fNL}(q_1,q_2,q_3,q_4,p)}{q_1q_2q_3q_4}\nonumber\\
& &\times F_{geom}(q_1,q_2,q_3,q_4,p),
\eea
with the geometric factor given by,
\bea
F_{geom}(q_1,q_2,q_3,q_4,p)
& \equiv & \int \de \tilde{\phi}_3\int \de \Omega_k~(\hat{\kv}\cdot\qv_1)(\hat{\kv}\cdot\qv_2)(\hat{\kv}\cdot\qv_3)(\hat{\kv}\cdot\qv_4)\nonumber\\
& = & 
\frac{1}{120}\left[(q_1^2-q_2^2)^2-7 (q_1^2+q_2^2)p^2+6p^4\right]
   \nonumber\\& &
   -\frac{1}{30} \left[(q_1^2-q_2^2)^2-3 (q_1^2+q_2^2)p^2 +2 p^4\right]\frac{\left(q_3^2+q_4^2\right)}{ p^2}
   \nonumber\\& &
    +\frac{1}{120}\left[3(q_1^2-q_2^2)^2-5(q_1^2+q_2^2)p^2+2p^4\right]\frac{\left(q_3^2-q_4^2\right)^2}{p^4},
\eea
being a function of the five variables. The $\gNL$ integral follows
from the same considerations, with
$\widetilde{T}_{\gNL}(q_1,q_2,q_3,q_4)$ replacing
$\widetilde{T}_{\fNL}(q_1,q_2,q_3,q_4,p)$. An explicit evaluation of
these integrals for the cosmology assumed in this paper (see
Sec.~\ref{sec:localmodel}) yields
\beq
\la(\vd\cdot\kv)^4\ra_{c}=k^4 \, \fNL^2 \, 9.4 \, \times10^{-6} + k^4
\, \gNL \, 3.1\times 10^{-7}
\eeq
assuming $T_0$ is evaluated at $z=0$ and $\vk$ is in units of $\kMpc$.

\end{document}